\documentclass[iop]{emulateapj}
\usepackage{amsmath,graphics,epsfig}
\usepackage{natbib,hyperref}
\usepackage{xcolor,soul}
\usepackage{enumerate}

\defcitealias{gallet2015}{GB15}
\defcitealias{sestito2005}{SR05}

\begin{document}

\title{Lithium depletion is a strong test of core-envelope recoupling}

\shorttitle{Li depletion from core-envelope recoupling}

\author{Garrett Somers \& Marc H. Pinsonneault}

\affil{Department of Astronomy, The Ohio State University\\
    140 West 18th Ave, Columbus OH 43210, USA\\
    somers@astronomy.ohio-state.edu}

\def\teff   {{$T_{\rm eff}$}}
\def\teffs  {{$T_{\rm eff}s$}}
\def\msun   {{M$_{\odot}$}}
\def\rsun   {{R$_{\odot}$}}
\def\ML     {{$\alpha_{\rm ML}$}}
\def\MLsol  {{$\alpha_{\odot}$}}
\def\fk     {{$f_k$}}
\def\fc     {{$f_c$}}
\def\wsat   {{$\omega_{\rm crit}$}}
\def\chired {{$\chi^{2}_{\rm{red}}$}}
\def\cms    {{$\rm{cm}^2\ \rm{s}^{-1}$}}
\def\do     {{$D_{0}$}}
\def\tauce  {{$\tau_{CE}$}}

\begin{abstract}

Rotational mixing is a prime candidate for explaining the gradual depletion of lithium from the photospheres of cool stars during the main sequence. However, previous mixing calculations have relied primarily on treatments of angular momentum transport in stellar interiors incompatible with solar and stellar data, in the sense that they overestimate internal differential rotation. Instead, recent studies suggest that stars are strongly differentially rotating at young ages, but approach solid body rotation during their lifetimes. We modify our rotating stellar evolution code to include an additional source of angular momentum transport, a necessary ingredient for explaining the open cluster rotation pattern, and examine the consequences for mixing. We confirm that core-envelope recoupling with a $\sim$20 Myr timescale is required to explain the evolution of the mean rotation pattern along the main sequence, and demonstrate that it also provides a more accurate description of the Li depletion pattern seen in open clusters. Recoupling produces a characteristic pattern of efficient mixing at early ages and little mixing at late ages, thus predicting a flattening of Li depletion at a few Gyr, in agreement with the observed late-time evolution. Using Li abundances, we argue that the timescale for core-envelope recoupling during the main sequence decreases sharply with increasing mass. We discuss implications of this finding for stellar physics, including the viability of gravity waves and magnetic fields as agents of angular momentum transport. We also raise the possibility of intrinsic differences in initial conditions in star clusters, using M67 as an example.

\end{abstract}

\section{Introduction}\label{sec1}

The light element lithium is a sensitive diagnostic of mixing in stars. This is a consequence of its fragility: Li is easily destroyed by proton capture in stellar interiors at temperatures $\sim 2.5 \times 10^6$~K. Because standard stellar models treat convection as the only source of mixing, they predict Li depletion on the pre-main sequence, when stars possess deep convective envelope with hot bases, but not on the \textit{main sequence} for solar-mass stars and hotter. However, open cluster studies have revealed that Li is depleted by factors of 10 or more during the main sequence \citep[][SR05 hereafter]{zappala1972,sestito2005}, and indeed our own Sun is depleted by $\sim 40 \times$ compared to expectations from standard standard models initialized on the solar meteoritic abundance scale \citep{iben1965,lodders2003}. Moreover, dispersions in Li at fixed mass are evident in the Li abundance patterns of many open clusters \citep[e.g.][]{soderblom1993a,thorburn1993,jones1999}, while standard models predict no such dispersion. This is a sure signature of the occurrence of non-standard mixing in solar-type stars.

These observational features can be interpreted as a consequence of rotationally-induced mixing pumping Li-depleted material from the deep interior into the convective envelope. This explanation is attractive, as the energetics of stellar rotation are sufficient to produce the required mixing, and the observed dispersion of natal rotation rates naturally predicts varying rates of mixing, and thus dispersions in Li at fixed age. Moreover, the rate of rotationally induced Li depletion is generically predicted to decline with age, consistent with the data. However, previous rotational mixing calculations have generally relied on incomplete treatments of internal angular momentum transport, which cannot reproduce the observed rotational evolution of open clusters. In this paper, we propose to incorporate into our mixing models a more sophisticated treatment of angular momentum transport, and determine the consequences for Li depletion.

Solar-type stars are born rapidly rotating ($P_{\rm rot} = 0.1-10$~d), and subsequently spin down by factors of 5-500 before exhausting their hydrogen cores. Substantial work has gone into understanding this spin down, and over the past decades a consistent picture has emerged. Mass driven from the surface by magnetized winds robs the outer convective zone of angular momentum \citep{weber1967}, causing the envelope to decouple from the radiative interior and spin down. This process produces strong differential rotation between the slowly-rotating envelope and the rapidly-rotating core. Once on the main sequence, transport mechanisms drive angular momentum out of the core into the envelope, leading the radiative and convective regions to converge in rotation over some timescale set both by underlying physics of the transport processes and the unique parameters of the individual star \citep{pinsonneault1989}. Thus, for both single stars and members of multiple systems which are adequately separated from their companions, two main effects govern angular momentum evolution, 1) loss at the surface, 2) transport between interior layers. 

The nature of these internal transport mechanisms is an important open question in stellar physics. The only securely-known processes are those resulting from hydrodynamic instabilities \citep[for an extensive review, see][]{talon2008}, such as meridional circulation \citep{eddington1929,sweet1950}, associated baroclinic instabilities \citep{zahn1992}, shear instabilities \citep{zahn1974,maeder1998}, and the Goldreich-Schubert-Fricke mechanism \citep{goldreich1967,fricke1968}. However, early rotation models incorporating solely hydrodynamic effects predicted strong differential rotation at the solar age \citep[e.g.][]{pinsonneault1989,chaboyer1995}, and could not explain the near-solid-body solar rotation profile inferred from helioseismology \citep{libbrecht1988,schou1998}. Furthermore, they could not simultaneously explain the evolution of the rapid and slow rotating branches of open clusters from the pre-main sequence to the solar age \citep{krishnamurthi1997}.

By contrast, recent rotation models have found success in predicting the evolution of open cluster rotation rates by supposing that angular momentum transport is in fact \textit{stronger} than the predictions of pure hydrodynamics \citep[e.g.][]{macgregor1991,denissenkov2010,gallet2013,lanzafame2015}. These papers have followed two general patterns. Some use the so-called ``two-zone model'', which treats the convective envelope and radiative core as solid body regions, whose rotation rate can differ from one another, but who converge in rotation with a specified core-envelope coupling timescale \tauce\ \citep[e.g.][]{macgregor1991,gallet2013}. Others assume the existence of a background source of angular momentum transport that operates in conjunction with hydrodynamics \citep{denissenkov2010}. As both internal magnetic fields \citep[e.g.][]{gough1998,eggenberger2005} and gravity waves \citep{press1981,ando1986,charbonnel2005} are known to operate in stellar interiors, they are considered strong candidates for the background mechanism. Both techniques conclusively demonstrate that one or more additional sources of internal angular momentum transport are required to explain the rotational evolution of stars.

For all their success, these recent papers have predominately focused on rotation, with less attention given to the impact of these additional transport processes on an important consequence of angular momentum evolution: rotational mixing \citep[e.g.][]{pinsonneault1997}. Rotational mixing occurs in concert with hydrodynamic angular momentum transport, since the hydrodynamic instabilities discussed above physically transport material from one layer to another, carrying the specific abundances of chemical species within the transported parcel of mass \citep{pinsonneault1989,chaboyer1995,li2014}. Rotational mixing can alter the observed chemical abundances of stars by mixing nuclear-processed material into the surface layers, and can extend the lifetimes of stars beyond the predictions of standard stellar theory by replenishing the nuclear core with fresh hydrogen. Core-envelope recoupling would naturally impact these processes, as it fundamentally alters the angular momentum history of stars. It is worth noting though that the additional transport mechanisms typically invoked (magnetic fields, gravity waves) do not \textit{directly} mix chemical material, but instead modulate the internal angular momentum profile, which in turn influences the rate of hydrodynamic mixing.

Given the important consequences of mixing, and the patent need for non-hydrodynamic angular momentum transport in stellar interiors, we investigate the impact of core-envelope recoupling on mixing. We do this by including in our fully-consistent evolution code a constant diffusion coefficient (\do), which increases the diffusion of angular momentum at all layers in the radiative region by a fixed amount \citep{denissenkov2010}, but does not mix chemical material (see $\S$\ref{sec2.3} for details). This approach is better suited to study mixing than the two-zone model, as Li depletion depends sensitively on the degree of differential rotation in the outer layers of the core, structure which two-zone calculations do not resolve.

We also introduce a new method for calibrating main sequence mixing models, which seeks to mollify certain inconsistencies between typical mixing calculations and their empirical constraints. The most egregious example is the spuriously strong Li depletion predicted by mixing models during the pre-main sequence: evolutionary models predict that solar-mass rapid rotators significantly deplete Li during their first 20~Myr, whereas the solar-mass stars in the Pleiades are no more Li-poor than their slowly rotating counterparts \citep[e.g.][]{ventura1998,somers2015b}. We address this issue by initializing our Li abundances on the zero-age main sequence. A second example are complications arising from interactions between young stars and their circum-stellar accretion disks, which have profound consequences for the angular momentum evolution of T Tauri stars. As such stars are the anchor for the initial rotation distribution of population models, the obfuscating effects of disk-locking must be dealt with. Finally, previous mixing exercises have typically used the Sun as the calibrator both for the rate of angular momentum loss, and for the strength of chemical mixing. However, pegging a physical model to a single point could be misleading, if that datum were an outlier. Calibrating instead to an ensemble average will be more robust, because a distribution of initial conditions produces a distribution of Li depletion factors. In order to produce informative models of main sequence Li depletion, we address each of these complications in our calibration procedure ($\S$\ref{sec3.4}).

We incorporate these techniques into our rotating evolution code, and explore the mixing signature arising from core-envelope recoupling driven by enhanced angular momentum transport. As we will show, these ``hybrid'' models, as we call them, can simultaneously replicate both the open cluster rotation and Li depletion pattern far better than other models of stellar evolution. Furthermore, we reveal a strong mass dependence in the rate of core-envelope coupling, which has been seen previous in rotation, and which we demonstrate for the first time using lithium.

Our paper is organized as follows. We describe our evolutionary models, and our adopted calibration data sets, in $\S$\ref{sec2}. We then introduce both the rotation and mixing behavior of our hybrid models, and compare/contrast them with traditional limiting cases of angular momentum transport ($\S$\ref{sec3}). In $\S$\ref{sec3.4}, we lay out our calibration scheme. We then calibrate the efficiency of core-envelope recoupling in solar-mass stars using open cluster rotation data ($\S$\ref{sec4.1}) and the efficiency of mixing in several mass ranges using open cluster Li data ($\S$\ref{sec4.2}). From these results, we derive the implied core-envelope coupling timescales ($\S$\ref{sec4.3}), and discuss implications for gravity waves and magnetic fields as transport agents. In $\S$\ref{sec5}, we compare our models to the internal rotation profile of the Sun ($\S$\ref{sec5.2}), and discuss the possible existence of cosmic variance in open cluster properties ($\S$\ref{sec5.3}). Finally, $\S$\ref{sec6} concludes.

\section{Methods}\label{sec2}

In $\S$\ref{sec4}, we constrain the rate of angular momentum loss, the required strength of our constant diffusion coefficient, and the efficiency of mixing in a range of stellar models. To do this, we employ spot-modulation rotation periods and Li abundances from a series of open clusters, and utilize our stellar evolution code.

\subsection{Rotation data}\label{sec2.1}

For this paper, we will compare our rotational models to statistical bounds of the open cluster rotation distribution as a function of age. While the entire distribution could in principle be modeled, this technique is far more sensitive to sample contamination, such as cluster non-member backgrounds, and binary components whose angular momentum evolution has been influenced by tidal forces from their companions. Instead, we will compare our models to the 25th (slow), 50th (median), and 90th (fast) percentiles of rotation rates within the open clusters. By taking these ``rotational moments''  of a full distribution, much of this contamination can be averaged out, and a realistic picture of the evolution of the distribution emerges.

Our rotation comparison data comprises the evolution of the slow, median, and fast moments of the $0.9-1.1$~\msun\ open cluster rotation distribution, as reported by \citet[][GB15 hereafter]{gallet2015}. For all but two of the clusters, we adopt the ages listed in table 1 of \citetalias{gallet2015}, and assume a 10\% age error, following the recommendation of \citet{soderblom2010}. The exceptions are the Hyades and Praesepe for which a recent study has reported an age of $\sim 800$~Myr \citep{brandt2015}, distinct from the $\sim 600$~Myr values adopted by \citetalias{gallet2015}; we adopt a compromise age of $700 \pm 100$~Myr. For additional constraints on the behavior at early ($< 100$~Myr) and late ($>1$~Gyr) ages, we further include the rotational moments of the 80 Myr $\alpha$ Persei and the 2.5 Gyr NGC 6819 \citep{irwin2009b,meibom2015}. 

To quantify the agreement between a given rotational track and the cluster data, we adopt a reduced $\chi^2$ metric, taking into account the age uncertainty of each cluster and the rotation errors on each cluster moment. For a given stellar model, we determined the closest approach of its rotational track to each error ellipse, in terms of the $\sigma$ offset. We then summed the square of the $\sigma$ values resulting from each data point, excluding those points employed during the calibration process, and divided the square root of this sum by the total number of data points to obtain a reduced $\chi^2$. When fitting multiple rotation moments simultaneously, we required the ``best-fit'' age determined for each cluster to be equal in the two panels.

This methodology contains the implicit assumption that open clusters form a perfect evolutionary sequence. That is, a young cluster like the Pleiades, given time, will evolve into the rotational distribution revealed by older clusters, like the Hyades and M67. This assumption may break down if there exists substantial variance in the initial conditions of open clusters, which could influence e.g. the fraction of rapid vs. slow rotators. Having made this assumption explicit, we will cautiously proceed, and return to the question in $\S$\ref{sec5.3}.

\subsection{Lithium data}\label{sec2.2}

For our mixing calibration, we choose as a comparison sample the data compilation of \citetalias{sestito2005}. These authors collected photometry and Li equivalent widths in open clusters available prior to 2005, grouped the clusters into age bins, and uniformly reanalyzed the \teffs\ and Li abundances of their members in order to study the evolution of Li with time. We collected their analyzed data from the WEBDA database, and for each of their specified age bins (see \citetalias{sestito2005}, table 3), computed the mean abundance within the \teff\ bin $5750 \pm 100$~K, for comparison with our solar mass models. We noted in \citet{somers2014} that their reported members for the 8~Gyr old NGC 188 cluster were in fact sub-giants, rendering them an invalid comparison for the young clusters; accordingly, we have excluded them from our analysis.

\subsection{Stellar models}\label{sec2.3}

We calculate stellar evolution models using the Yale Rotating Evolution Code (YREC). We adopt a present day solar metal abundance of Z/X = 0.02292 \citep{grevesse1998}, accounting for helium and metal diffusion \citep{thoul1994}, and calibrate the hydrogen abundance and mixing length such that a 1.0\msun\ stellar model achieves the present-day solar luminosity of $3.8418 \times 10^{33}$~erg~s$^{-1}$ and solar radius of $6.9591 \times 10^{10}$~cm at the solar age of $4.568$~Gyr. Our models use nuclear cross-sections from \citet{adelberger2011}, model atmospheres from \citet{castelli2004}, equation of state tables from OPAL 2006 \citep{rogers1996,rogers2002}, high and low temperature opacities from \citet{mendoza2007} and from \citet{ferguson2005} respectively, and the electron screening treatment of \citet{salpeter1954}.

To model stellar winds, we use a Rossby-scaled version of the the wind law of \citet{chaboyer1995}, a modification of the original prescription of \citet{kawaler1988}. This formulation includes three free parameters which must be calibrated: the normalization \fk, the saturation threshold \wsat, and a measure of the magnetic field geometry $N$. Angular momentum loss obeys the formula,

\begin{equation} 
 \displaystyle \frac{dJ}{dt} = \left\{
	\begin{array}{l l}
	 \displaystyle f_K \left(\frac{R}{R_{\odot}}\right)^{N-1} \left(\frac{M}{M_{\odot}}\right)^{-N/3} \omega \omega_{crit}^{4N/3}, \\
	 
	 \hspace{4cm} \omega > \omega_{crit} \frac{\tau_{CZ,\odot}}{\tau_{CZ}} \\ \\
	 
	 \displaystyle f_K \left(\frac{R}{R_{\odot}}\right)^{N-1} \left(\frac{M}{M_{\odot}}\right)^{-N/3} \omega \left( \frac{\omega \tau_{CZ}}{ \tau_{CZ,\odot}} \right)^{4N/3}, \\
	 
	 \hspace{4cm} \omega \leq \omega_{crit} \frac{\tau_{CZ,\odot}}{\tau_{CZ}} \\
	 
	\end{array} \right. 
	\label{eqn:kawaler}
\end{equation}

where $M$ and $R$ are the stellar mass and radius, $\omega$ is the rotation frequency, and $\tau_{CZ}$ is the global convective overturn timescale as defined in \citet{vansaders2013}.

As we are modeling the slow, median, and fast bounds of the open cluster distribution, we require three separate initial rotation conditions. We choose initial rotation periods and disk-locking lifetimes which reproduce the slow, median, and fast bounds of the 13 Myr h~Persei open cluster  \citep[for details, see][]{somers2015b}. The three free parameters in the wind law are calibrated such that the median rotating model reproduces the median rotation rate of M37 at 550~Myr and the solar rotation rate at 4.568~Gyr, and such that the fast rotating model matches the 90th percentile rotators in the Pleiades. In the case of hydro models (see $\S$\ref{sec3.1}) we relax the M37 constraint, as we find they are intrinsically unable to predict both M37 and the Sun with a single wind law ($\S$\ref{sec4.1}).

Internal angular momentum transport and chemical diffusion obey the coupled differential equations described by \citet{pinsonneault1989},

\begin{equation} 
\frac{\rho r^2}{M} \frac{d J}{dt} = f_{\omega} \frac{d}{dr} \left( \frac{\rho r^2}{M} (D_{\rm hydro} + D_{0}) I \frac{d\omega}{dr} \right),
\label{eqn:amt1}
\end{equation}

\begin{equation}
\rho r^2 \frac{d X_i}{dt} = f_c f_{\omega} \frac{d}{dr} \left( \rho r^2 D_{\rm hydro} \frac{dX_i}{dr} \right).
\label{eqn:amt2}
\end{equation}

Here, $I$ is the moment of inertia, $M$ and $J$ the mass and angular momentum of the given shell, $D_{\rm hydro}$ is the sum of diffusion coefficients of the various hydrodynamic transport processes mentioned in $\S$\ref{sec1} \citep{pinsonneault1989,pinsonneault1991,chaboyer1995}, $f_\omega$ is a scaling factor that sets the efficiency of this transport ($f_\omega$ = 1 in our models), and \fc\ is a scaling factor that sets the efficiency of transport of chemical species $X_i$ relative to $f_\omega$, related to anisotropic turbulence \citep{chaboyer1992}. Our numerical scheme iteratively solves for the $I$ and $\omega$ which correspond to $J$ at each shell, and implicitly conserves angular momentum by considering the impact of rotation on the moment of inertia.

These details follow our previous work, but in this paper we introduce a new feature of our evolution code, a constant diffusion coefficient \do\ \citep[e.g.][]{denissenkov2010}. This value is a free parameter of our models, and enhances the diffusion of angular momentum by the specified value at every location in the radiative zone, but not the convective zone. This diffusion influences only angular momentum, and does not transport chemicals, and thus represents a background source of non-hydrodynamic transport. Its influence on stellar rotation and mixing are explored in detail in the following section.

\section{Rotation and mixing in stellar models}\label{sec3}

In this section, we introduce the generic behavior of our rotating evolutionary models. First, we describe the rotational and mixing properties of ``hydro'' and ``solid body'' models. These are referred to collectively as the limiting cases, as they represent, respectively, the \textit{weakest and strongest} possible coupling of the angular momentum histories of adjacent layers in the interiors of stars. Then, we use the limiting cases as a framework for introducing our hybrid models, and describing how angular momentum transport in stellar interiors influences surface rotation and internal mixing.

\begin{figure*}[t]
\begin{centering}
\includegraphics[width=6.5in]{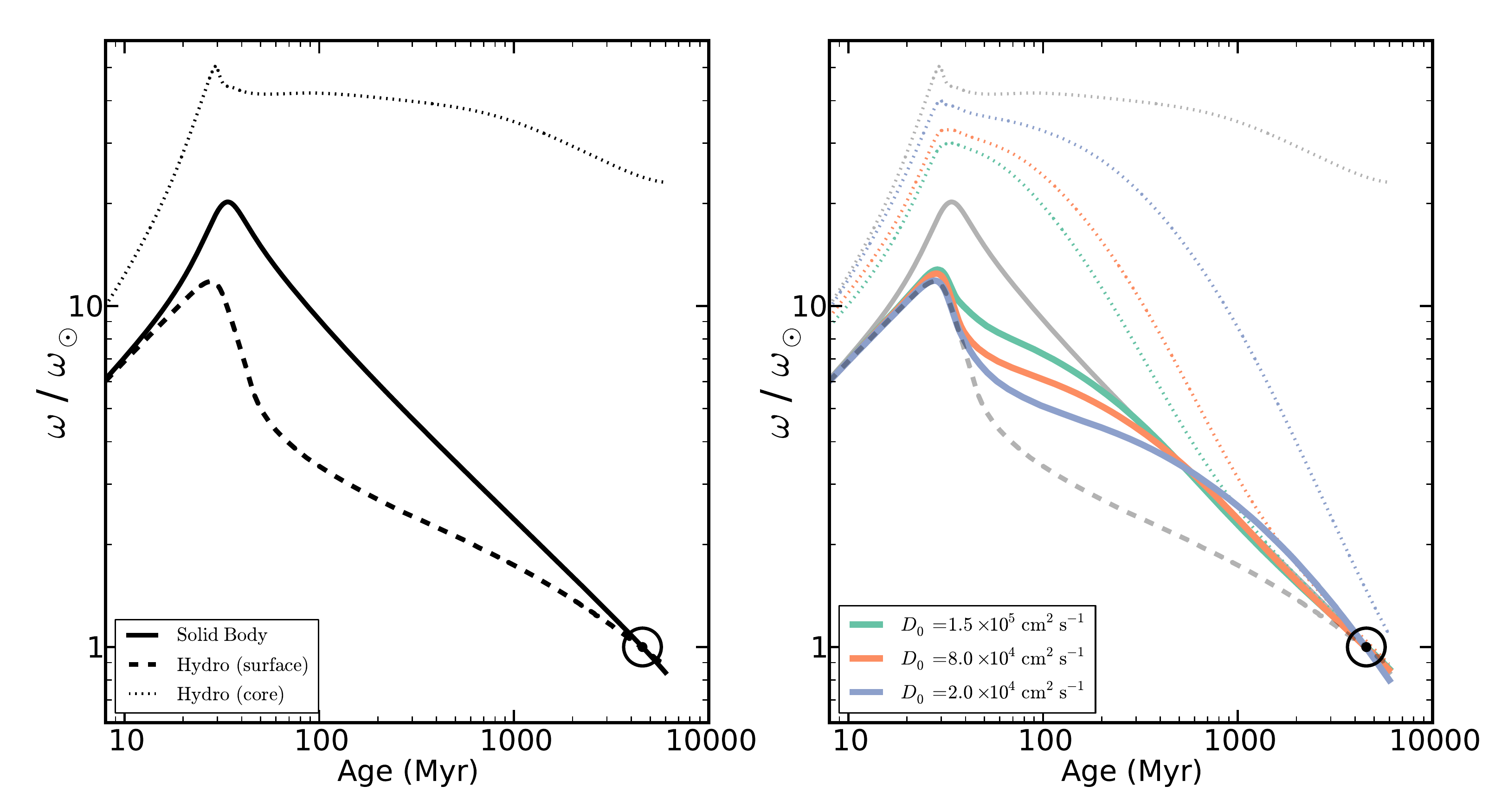}
\caption{The surface and core rotational evolution of solar mass models, employing various internal angular momentum transport paradigms. The models include the surface and core of the solid-body model (solid black), the hydro model (dashed and dotted black), and hybrid models with three different values of the constant diffusion coefficient in blue, orange, and green. Each model's wind law has been calibrated to reach the solar rotation rate at the solar age.}
\label{fig1}
\end{centering}
\end{figure*}

\subsection{Rotation in limiting case models}\label{sec3.1}

In hydro models, internal angular momentum transport is driven by structural evolution (particularly on the pre-main sequence) and by a number of hydrodynamic instabilities (particularly on the main sequence; see \citealt{pinsonneault1997}), which generally depend on the local rotation rate (e.g. meridional circulation), and/or the local rotation \textit{gradient} (e.g. shear instabilities). Hydro models consider only these instabilities, locally conserving angular momentum otherwise, and thus represent the weakest possible coupling of a stellar interior permitted by hydrodynamics. Note that in the presence of strong horizontal turbulence, hydrodynamic transport is an inherently diffusive process, so angular momentum will preferentially move from layers that are rapidly rotating to layers which are slowly rotating \citep{chaboyer1992,zahn1992}. This will generally result in outward diffusion in our evolutionary models, as the only angular momentum sink is the magnetized wind at the surface. 

The rotational evolution of the surface and core of a 1\msun, hydro model are shown by the dashed and dotted lines in Fig. 1. Significant differential rotation has developed during the pre-main sequence, following the emergence of a radiative core. After entrance onto the main sequence ($\sim 30$~Myr), the surface begins to rapidly spin down while the core remains nearly constant in its rotation. By $\sim 100$~Myr, the internal differential rotation is so strong that the hydrodynamic transport processes which scale with the rotation gradient become influential. Diffusion of angular momentum from the core to the envelope slows the spindown rate of the surface by partially counter-balancing the loss at the surface, and the radiative core begins to gradually slow down. The surface reaches the solar rotation rate at 4.6~Gyr, but the core is still rotating $\sim 20 \times$ faster, strongly inconsistent with the near-solid-body behavior exhibited by the Sun \citep[e.g.][]{eff-darwich2002,couvidat2003}.

By contrast, solid body models force each layer in the radiative core to rotate with the same period as the convective envelope. Thus, angular momentum redistribution can be thought of as extremely efficient in such models, as any loss from the surface is instantly balanced by transport from the interior. The rotational evolution of both the surface and core of a 1\msun, solid body model is shown by the solid black line in Fig. 1. During the first tens of Myrs, the surface rotation gradually increases due to Hayashi contraction. The peak rotation rate achieved at the zero-age main sequence is higher for the solid body case, because the lost surface angular momentum is instantaneously replenished, compared to the hundreds of Myrs required in hydro models. Once on the main sequence, the star spins down at a linear rate due to magnetized stellar winds, finally arriving at the solar rotation rate at 4.6~Gyr with no internal differential rotation.

\subsection{Rotation in hybrid models}\label{sec3.2}

We now introduce our hybrid models which, in addition to the angular momentum diffusion induced by hydrodynamics, drive additional diffusion, the strength of which is constant in time but a free parameter. As this additional transport operates over a set timescale, its influence will be minimal at early times, but increasingly important with age. Note that hydro and solid body models are in fact special cases of hybrid models, where \do\ is set to zero (no additional transport beyond hydrodynamics) or infinity (instantaneous coupling), respectively.

The rotational evolution of hybrid models with three different choices of \do\ are shown in the right panel of Fig. 1\footnote{The three choices of \do\ represented in this plot, $1.5 \times 10^5$, $8 \times 10^4$, and $2 \times 10^4$ \cms, respectively correspond to approximate core-envelope coupling timescales of 15, 25, and 85 Myr for a 1\msun\ star.}. These hybrid models mimic hydro models in their surface rotation rates on the early main sequence, as insufficient time has passed for the constant diffusion coefficient to exert much influence. However, their core and surfaces converge in rotation rate over a much shorter time period due to the enhanced transport. This flux of angular momentum from the core to surface initially boosts the surface rotation rate relative to the hydro case around $\sim 50$~Myr, and ultimately leads to solid-body-like behavior after a few hundred Myrs. The age at which this transition occurs depends on \do, with a stronger flux leading to more rapid core-envelope coupling. The strength of internal differential rotation at the solar age is also a strong function of \do, but can reach essentially solid-body behavior for \do\ $\gtrsim 5 \times 10^4$ \cms\ ($\S$\ref{sec5.2}).

The basic picture that emerges from our considerations are that hybrid models resemble hydrodynamic models when young (highly decoupled), and resemble solid body models when old (highly coupled). This will prove important when we consider the mixing signal arising from our three transport paradigms.

\subsection{Mixing}\label{sec3.3}

\begin{figure*}[t]
\begin{centering}
\includegraphics[width=6.5in]{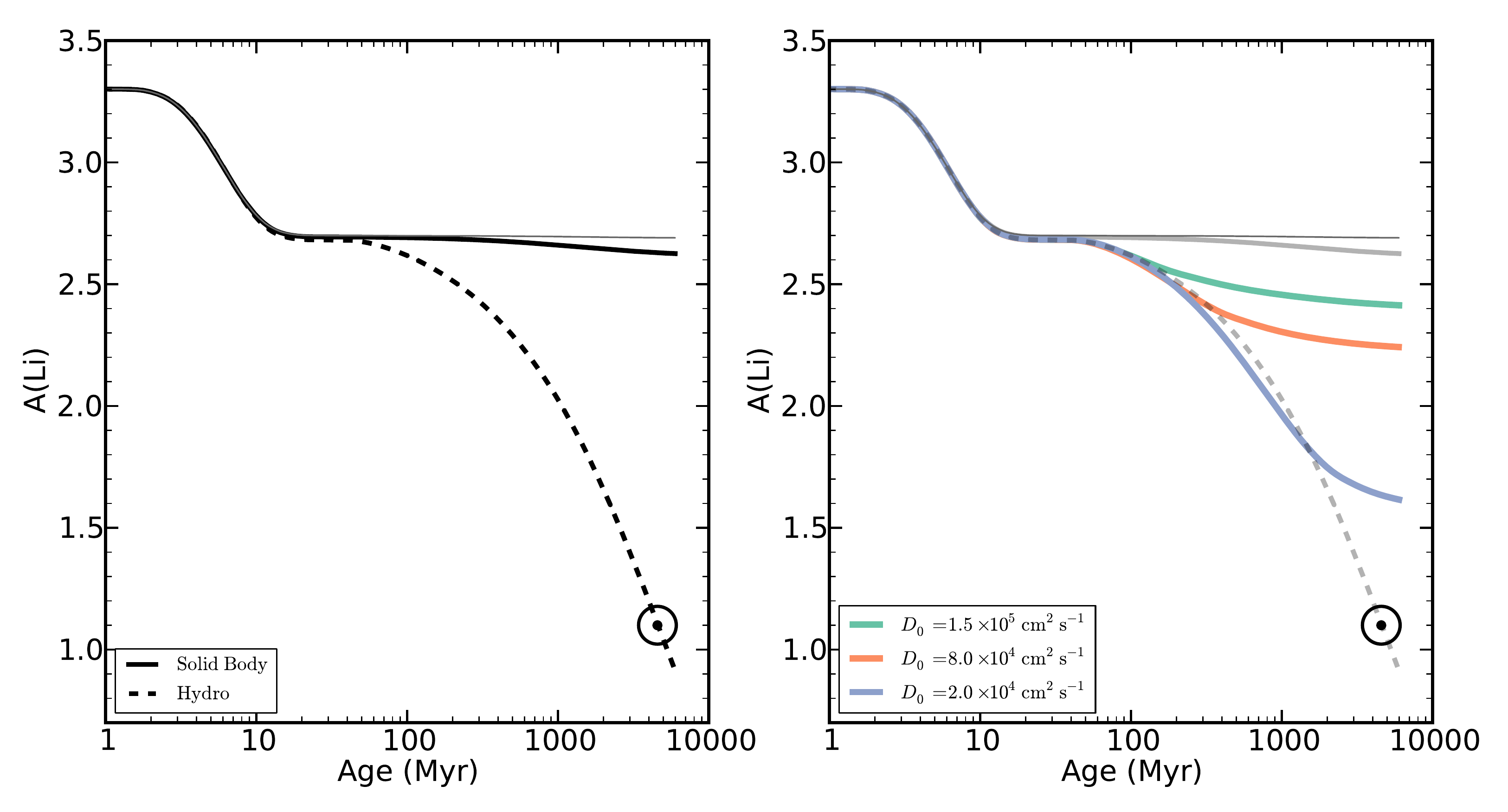}
\caption{Surface Li depletion for the solid-body (solid black), hydro (dashed black), and hybrid (colored) models of Fig. \ref{fig1}. The mixing parameter \fc\ has been calibrated such that the hydro model attains the Li abundance of the Sun at the solar age, and the other models are shown with the same mixing efficiency for comparison. Stronger core-envelope coupling leads to less efficient mixing at late ages, causing hybrid models to flatten their Li curves after several hundred Myrs. The thin grey line represents a non-rotating model.}
\label{fig2}
\end{centering}
\end{figure*}

Mixing describes the transportation of chemical material between layers in the interior of stars. Hydrodynamic angular momentum transport is the vector for rotationally-driven mixing, and thus its history strongly influences the mixing pattern. Mixing has many observational tracers, but the most powerful is the surface abundance of lithium, which decreases when deep depleted material is mixed into the convective envelope, at a rate proportional to the mixing efficiency.

We set the initial Li abundance of our models to the primordial value of the Sun (A(Li)\footnote{$\rm{A(Li)} \equiv \log(\rm{Li/H}) + 12$ }~$\sim$~3.3), and present the evolution of the surface abundance predicted by the hydro and solid body models of $\S$\ref{sec3.1} in the left panel of Fig. \ref{fig2}. These are compared to a standard stellar model which does not rotationally mix, shown by the thin grey line. The efficiency of mixing relative to angular momentum transport, a free parameter of our model (\fc), is fixed to reproduce the solar abundance in the hydro model. We hold this fixed for the other models in this figure, in order to illustrate the relative mixing strengths of each transport scenario. The behavior differs in individually calibrated models ($\S$\ref{sec4.2}).

Each model undergoes a similar period of pre-main sequence depletion from $\sim 2 - 10$~Myr, when the surface convection zone is deep. After this period, the rate of Li depletion differs strongly between the two cases. The lack of significant differential rotation in the solid body model leads to very little hydrodynamic transport, and thus a very weak dilution of the surface Li abundances during the lifetime of the Sun. By contrast, the hydro model develops extremely strong internal shears (Fig. \ref{fig1}), and drives abundant transport though hydrodynamics, thus leaving a far larger imprint on the surface Li abundance. Somewhat paradoxically, stronger coupling leads to less efficient mixing, as the strong differential rotation which so efficiently mixes chemical material is never allowed to develop in solid body models. 

The right panel of Fig. 2 now shows the mixing signal derived from the three hybrid models of $\S$\ref{sec3.2}. Analogous to the rotation rates, the mixing signal of hybrid models follows a similar trajectory during the early years as do the hydro models (up to $\sim 100$~Myr). Thereafter, the internal shears become progressively suppressed by the enhanced angular momentum flow, reducing the efficiency of hydrodynamic transport and mixing. This leads to a gradual flattening of the Li depletion track. Thus, hybrid models are efficient mixers during their early years, and inefficient mixers at late ages. As before, the age of this transition is a strong function of \do, such that stronger transport leads to an early transition from efficient to inefficient, and thus a higher ultimate Li abundance.

We note that this late-time flattening is a generic prediction of models which are strongly differentially rotating on the early main sequence, and recouple over time, as it reflects a steady suppression of the shears in the interiors. Thus, even if our constant diffusion coefficient methodology does not capture the minutiae of internal angular momentum transport, the suppression of Li depletion at late ages should occur in any models which attempt to reproduce the flat solar rotation profile.

\section{Our calibration procedure}\label{sec3.4}

\begin{figure*}[t]
\begin{centering}
\includegraphics[width=7.0in]{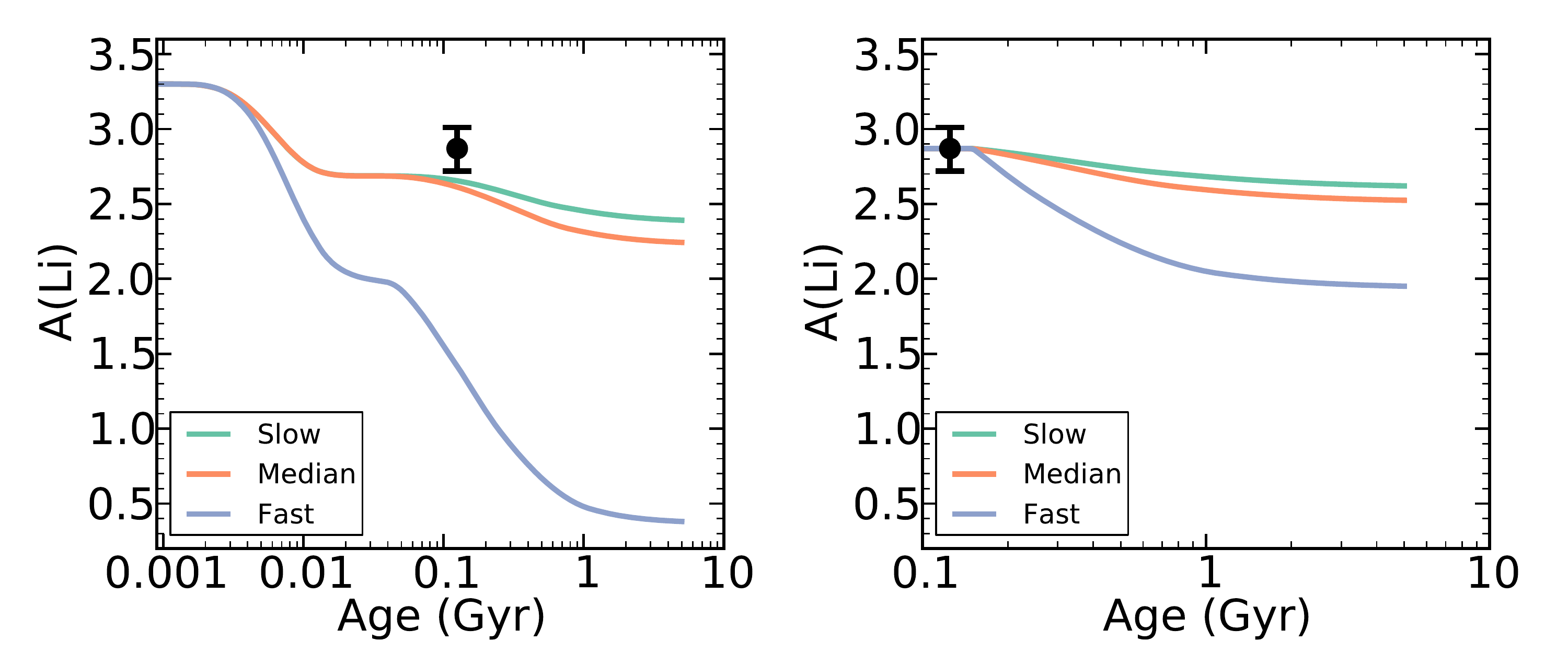}
\caption{An illustration of our pre-main sequence detrending process. \textit{Left:} The Li abundance tracks of three hybrid models of varying initial rotation rate. Li depletion is a strong function of rotation rate in the models, but this is contradicted by Li data from the Pleiades and Blanco~1 (black point), and explained by the structural effects of starspots \citep{somers2015a}. \textit{Right:} To account for this early failure, we reset the Li abundance of our tracks to the Pleiades abundance at 150~Myr, and track the depletion thereafter.}
\label{fig3}
\end{centering}
\end{figure*}

With the generic behavior of hybrid models laid out in $\S$\ref{sec3}, we now turn to our procedure for comparing these models to main sequence data. In $\S$\ref{sec1}, we briefly introduced the basic challenges, which include complications arising from accretion disk interactions, the pre-main sequence structural effects of magnetic fields, and uncertainties inherent to the solar calibration for both rotation and mixing. In this section, we detail our procedure for addressing these concerns.

First, during their formative years, protostars are surrounded by accretion disks. Proto-stars rotate well below break-up, indicating that angular momentum is efficiently extracted by interactions with these disks \citep{vogel1981,shu1994}. These stars would be expected to spin up dramatically as they contract, but there is a large population of slowly rotating stars, requiring a mechanism to spin down some systems early, but not all. Star-disk interactions are a natural explanation \citep{konigl1991}. A rotation distribution initialized during this ``disk-locking'' phase presents a poor initial launch point for models of wind-driven angular momentum loss. To circumvent this issue, we initialize our rotation rates on the 13~Myr old h~Persei cluster ($\S$\ref{sec2.1}). This cluster is older than the canonical disk-lifetime upper limit of 6--10~Myr \citep[e.g.][]{fedele2010}. Thereafter, one can be assured that magnetized winds are the dominant mechanism robbing the star of angular momentum.

Second, we must account for the strong discrepancy between rotational mixing models on the pre-main sequence, and the empirical pattern. The same structural effects that permit Li burning in the surface convection zone during the pre-main sequence make models highly sensitive to early rotationally-induced mixing. For this reason, mixing models tend to predict stronger pre-main sequence Li depletion with faster rotation \citep[e.g.][]{pinsonneault1989,charbonnel2005,somers2015b}. While this correlation seems intuitive, the Li patterns of young ($\lesssim 150$~Myr) open clusters indicate essentially no correlation between rotation and Li depletion for solar-mass stars\footnote{A strong correlation does set in towards lower masses, but the sign is opposite of the rotational mixing prediction: faster rotating stars are \textit{more Li-rich} than slow rotators!}. We illustrate this discrepancy with 1\msun\ hybrid models in the left panel of Fig. \ref{fig3}. The range in initial rotation rates leads to a large dispersion by 125~Myr, in sharp contrast with the solar-temperature Li dispersion found in the $\sim 125$~Myr \citep{stauffer1998} Pleiades and Blanco~1 open clusters (black error bar). Evidently, rotation has a smaller impact on pre-main sequence Li depletion in 1\msun\ stars than predicted by mixing models.

In our recent papers \citep{somers2015a,somers2015b}, we suggested that this effect is a consequence of a correlation between rotation rate and radius inflation, likely driven by the enhanced magnetic activity and spot coverage of the fastest spinning cluster members. As a result, the rapid rotators are inflated during the pre-main sequence, cooling their interior and suppressing the rate of Li depletion, both through direct destruction in the convection zone and through rotationally-induced depletion. Consequently, the fast and slow rotators have similar abundances at the zero-age main sequence. This suggestion is corroborated by the Li-rotation correlation present in cooler stars \citep{soderblom1993a,bouvier2016}, which is likely produced through structural changes to the fastest rotators \citep{somers2015b}. 

Fortunately, as stars approach the main sequence, their convection zone retreats and Li depletion becomes less sensitive to radius inflation. Thus, Li depletion can proceed as predicted by mixing models thereafter. As we are concerned in this paper solely with Li depletion arising on the main sequence, we can simply remove the complications influencing the early depletion signal by adopting a main sequence starting point. We therefore reset the Li abundances of our models to the median Li abundance of the Pleiades/Blanco~1 data set at 125~Myr, and track the evolution of Li thereafter. This technique is illustrated in the right panel of Fig. \ref{fig3}. In practice we find little sensitivity to the precise age adopted, due to our freedom to calibrate the mixing strength, as long as it is in the range of 100--200~Myr. With our models initialized on the main sequence, we have avoided the complicated physics of the pre-main sequence, and can safely trace the depletion and dispersion developing thereafter.

Finally, the traditional calibration procedure for mixing calculations involved reproducing the solar rotation rate and the solar Li abundance with a median rotator by tweaking the free parameters controlling wind loss and internal chemical mixing. While a useful strategy, two strong disadvantages of this approach are the lack of constraints on stellar behavior at earlier epochs, and the possibility that the Sun is atypical in its rotation and Li abundance. After all, there is no reason the Sun has to be a median star in a broad distribution of initial conditions. To address these concerns for the rotational calibration, we use the 550~Myr M37 as a joint calibration point alongside the Sun ($\S$\ref{sec2.1}). For the mixing calibration, we compute the average abundance of a series of open clusters of approximate age $\sim$ 600~Myr (the Hyades, Praesepe, Coma Ber, and NGC 6633; \citetalias{sestito2005}), and calibrate mixing to a median age anchor. This has the advantage of marginalizing over many stars with the goal of finding a more stable median, and of making the late-age mixing a pure prediction of the physics. We will show in $\S$\ref{sec4.2} that the evolution of Li abundances at ages $\gtrsim$~2~Gyr provide an excellent test of core-envelope recoupling, so an earlier calibration point is beneficial.

These steps, alongside our hybrid models, constitute the theoretical and methodological advances presented in this paper. In the subsequent section, we apply these principles to understand the angular momentum evolution and mixing history of solar-type stars.

\section{Results}\label{sec4}

\subsection{Constraining \do\ with rotation data}\label{sec4.1}

Having introduced the generic behavior of our hybrid models, we next appeal to open cluster data to discriminate between the three transport paradigms presented above. We will show that hybrid models provide a superior description of the open cluster rotation pattern, and derive a best-fit value for \do. This exercise will solidify a nominal rate of angular momentum transport for solar mass stars, thus allowing us to examine mixing in $\S$\ref{sec4.2}.

\begin{figure*}[t]
\begin{centering}
\includegraphics[width=7.0in]{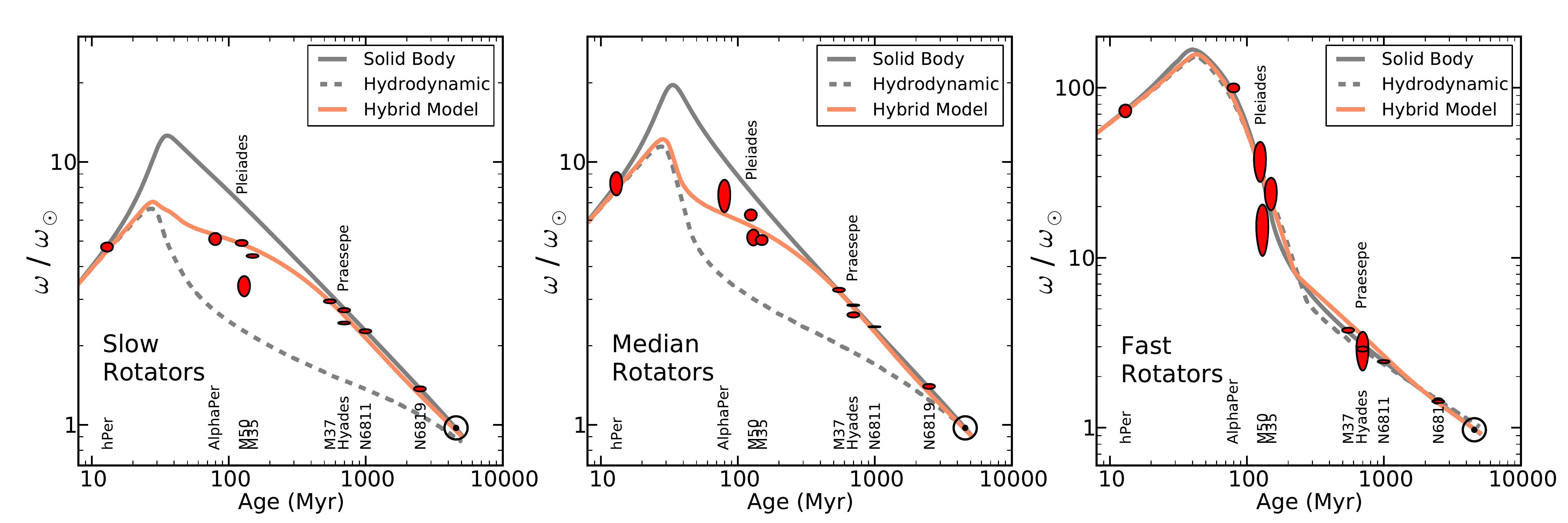}
\caption{A comparison of the surface rotation rates of our calibrated solid body, hydrodynamic, and hybrid models (calculated with \do\ = $9 \times 10^4$~\cms) and the rotation moments of \citetalias{gallet2015}, plotted as red 1$\sigma$ error ellipses. \textit{Left:} Solid body models do not spin down enough at early ages, and hydrodynamic models spin down too much at early ages, to explain the slowly rotating data. However, hybrid models spin down the right amount, and converge at later ages, producing better description of the data. \textit{Center:} Similarly, hybrid models out-perform the two limiting cases for describing the median rotators. \textit{Right:} Each rotation case performs reasonably similarly for the rapid rotators, owing to the strong hydrodynamic transport driven by rapid rotation. }
\label{fig4}
\end{centering}
\end{figure*}

\subsubsection{Limiting Cases}\label{sec4.1.1}
Fig. \ref{fig4} compares our evolutionary models with the rotation moments of \citetalias{gallet2015} (see $\S$\ref{sec2.1}), represented by red ellipses. Considering first the solid body models in solid grey, we find good agreement in the slow (left) and median (center) panels between the older clusters ($\gtrsim$~$500$~Myr) and the solar-calibrated rotation track. However, the solid body models cannot spin down enough at early times to reproduce the slow and median rotators in the young clusters ($\lesssim 500$~Myr). Turning to the hydro models (dashed grey), we find that the solar-calibrated track spins down \textit{far~too~much} to reproduce the slow and median bounds. Furthermore, hydro models cannot reproduce either the slope or normalization of the spindown rate implied by the older clusters. Suggestively, the young cluster data appears in between these two limiting cases. Considering finally the most rapid rotators, we see that both the solid body and hydro models trace similar paths. This is likely because hydrodynamic transport scales very strongly with absolute rotation rate, so that in the fastest case, sufficient angular momentum is diffused even in hydro models to counteract the winds at the surface. Both cases predict reasonably well the evolution of the upper envelope of rotation rates, with the exception of M34 (see below). Despite this, both the solid body and hydro models fail spectacularly to match the data (\chired\ = 32 and 407, respectively), owing to their failures in the two slower rotation bins.

\subsubsection{Hybrid Models}\label{sec4.1.2}

We next compare our hybrid models to the data. To do this, we varied the value of \do\ until the best-fit \chired\ was obtained for the median rotator bin. The resulting best-fit \do\ = $9 \times 10^4$ \cms. Hybrid models with this value appear in orange in Fig. \ref{fig4}. These models provide a far better qualitative agreement with the data, as their early (but not complete) decoupling reproduces the initial spin-down of the slow and median rotators younger than $\sim 500$~Myr, and their steady angular momentum flux gradually drives the model towards the solid body track, thus matching the data beyond 500~Myr. The conclusion from these panels is clear: hybrid models significantly outperform the adopted limiting cases, and demonstrate the need for core-envelope decoupling and recoupling during the main sequence.

The goodness-of-fit for the median bin is an excellent \chired\ = 1.04, demonstrating quantitatively the superiority of hybrid models. However, we find the \chired\ in the slow and fast bins to be 3.99 and 2.04, respectively. We identify two factors which may contribute to the poorness-of-fit. First, our \chired\ minimization exercise did not allow for a rotation dependence in \do. To check this, we determined the best fit \do\ for the slow and fast bins independently. We find that for the slow rotators alone, the best-fit \do\ of $6.5 \times 10^4$~\cms\ produces the improved \chired\ = 2.88, and for the fast rotators alone, \do\ of $1.3 \times 10^5$~\cms\ gives \chired\ = 1.47. These results hint that a rotation dependence in the rate of the additional angular momentum transport may be present. However, we caution that statistical noise from the next effect may be responsible.

Second, the rotation moments from some clusters in the \citetalias{gallet2015} sample are statistically inconsistent with one another. The most notable is M50, whose slow rotation moment of $3.38 \pm 0.30$~$\omega_{\odot}$ is quite discrepant from the equal-age Pleiades at $4.92 \pm 0.13$~$\omega_{\odot}$. This disagreement is likely a consequence of the high level of background-contamination in the M50 sample, estimated by the authors at 47\% \citep{irwin2009a}. As the majority of background contaminants will be old, and thus slowly rotating, we would expect an artificially low value for the 25th percentile. Excluding M50 brings the best-fit slow rotating model to \chired\ = 1.62, a significant improvement. Furthermore, we cannot exclude similar contamination in the other data sets, or true variance in the rotation distribution between open clusters \citep[e.g.][]{coker2016}, which would further disrupt comparisons between our models and the data. For the fast rotating bin, a similar role is played by the anomalously rapid 90th percentile in M34. This could result from the influence of synchronized binaries, a more dominant rapid rotator population in this particular cluster (see $\S$\ref{sec5.3}), or may simply be shot noise. Excluding this datum produces \chired\ = 1.00 when \do\ = $1.3 \times 10^5$~\cms\ is adopted.

To summarize, while the limiting case models are strongly ruled out, hybrid models are able to reproduce the qualitative and quantitative evolution of the slow, median, and fast rotating moments of open clusters extremely well. 

\subsection{Lithium depletion}\label{sec4.2}

We now consider the mixing signal arising from hybrid models with the best-fit \do\ = $9 \times 10^4$~\cms, and compare their results to cluster Li data from \citetalias{sestito2005} (see $\S$\ref{sec2.2}) as a function of age. We fix the mixing efficiency \fc\ such that the median age Li bin containing the Hyades, Praesepe, Coma~Ber, and NGC~6633 is reproduced, and examine the predictions for other age bins. We further apply our calibrated mixing efficiency to other \teff\ bins, in order to search for a mass dependence in the rate of core-envelope recoupling.

\begin{figure}[t]
\begin{centering}
\includegraphics[width=3.5in]{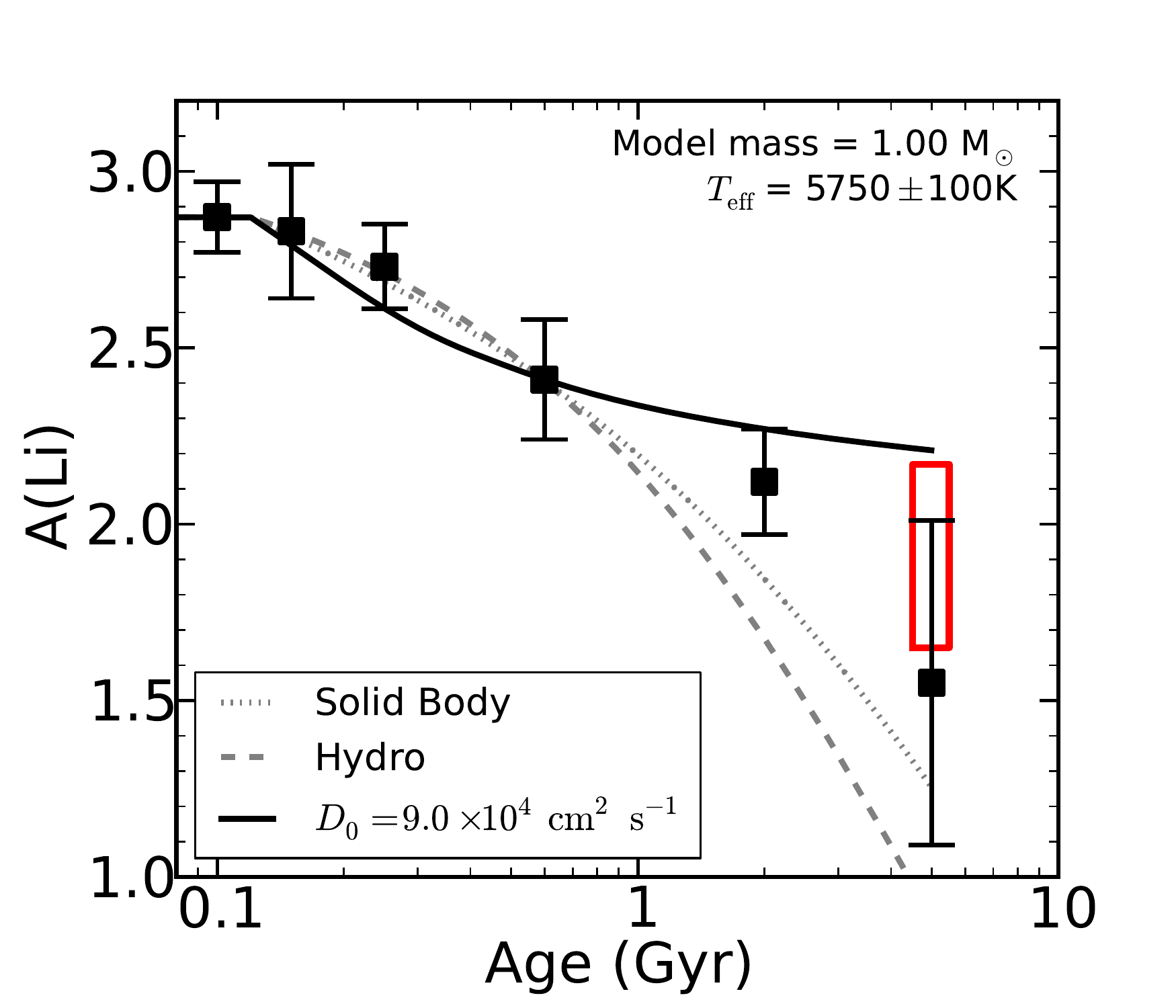}
\caption{Li abundances as a function of age (black squares) in the \teff\ range $5750 \pm 100$~K from \citetalias{sestito2005}, compared to solar mass models computed with our three angular momentum transport paradigms. The mixing efficiency \fc\ is calibrated for each model such that the data point at 600~Myr is reached. The two limiting cases predict a turn-down at late ages, while the hybrid models flatten. The final cluster, M67, appears to buck the prevailing trend, but the upper third of the data (red rectangle) matches the trend better.}
\label{fig5}
\end{centering}
\end{figure}

\subsubsection{Solar Mass Results}\label{sec4.2.1}

Fig. \ref{fig5} presents our solar mass results. The black squares denote the mean Li abundance of the \teff\ = $5750 \pm 100$~K Li bin as a function of age, and their error bars indicate the standard deviation of the scatter. The red rectangle at 5~Gyr represents the median and standard deviation of the \textit{top half} of the M67 data. The average abundance of the open clusters remain approximately flat until $\sim 200$~Myr, when they begin to gradually decrease. By 2~Gyr, the average star has depleted $\sim 0.7$~dex of Li. The final age bin, which consists solely of M67, is both substantially more depleted, and contains a far larger scatter, compared to the pattern suggested by the younger clusters. We discuss possible reasons for this in $\S$\ref{sec5.3}, but note that the top half of the M67 distribution (red rectangle) coincides well with the depletion trend of the younger clusters.

The three lines show Li depletion tracks for solar-mass models calculated for the two limiting cases, and a hybrid model computed with the best-fit, median-rotating, solar-mass \do\ of $9 \times 10^4$~\cms. The mixing efficiency \fc\ has been set for each model so that the Li abundance of the 600~Myr data point is accurately predicted; resulting values are 0.08, 0.21, and 9.50 for the hydro, hybrid, and solid body models, respectively. The difference between this result and Fig. \ref{fig2} reflects the impact of calibrating the model Li depletion to a benchmark system. The limiting cases are shown simply for comparison, as their incompatibility with the rotation evolution of open clusters has already been demonstrated in $\S$\ref{sec4.1}. The hybrid model, on the other hand, was favored by $\S$\ref{sec4.1}, and agrees well with the shape of the trend up to 2~Gyr. It cannot account for the sudden drop off of the mean of M67, but does coincide with the upper envelope of the M67 distribution. 

The visual success of the hybrid predictions for Li depletion in Fig. \ref{fig5} are driven entirely by calibrating the free parameter \fc, and are therefore not a compelling argument on their own. However, previous authors such as \citet{denissenkov2010} and \citet{lanzafame2015} have found that the timescale for core-envelope coupling \textit{decreases with increasing mass}. A fixed mixing efficiency therefore allows us to search for a mass dependence in the coupling rate implied by mixing. Thus, we expect that a fixed \do\ with mass produces poor agreement with the data, but an increasing \do\ with mass will provide a better description.

\begin{figure}
\begin{centering}
\includegraphics[width=3.5in]{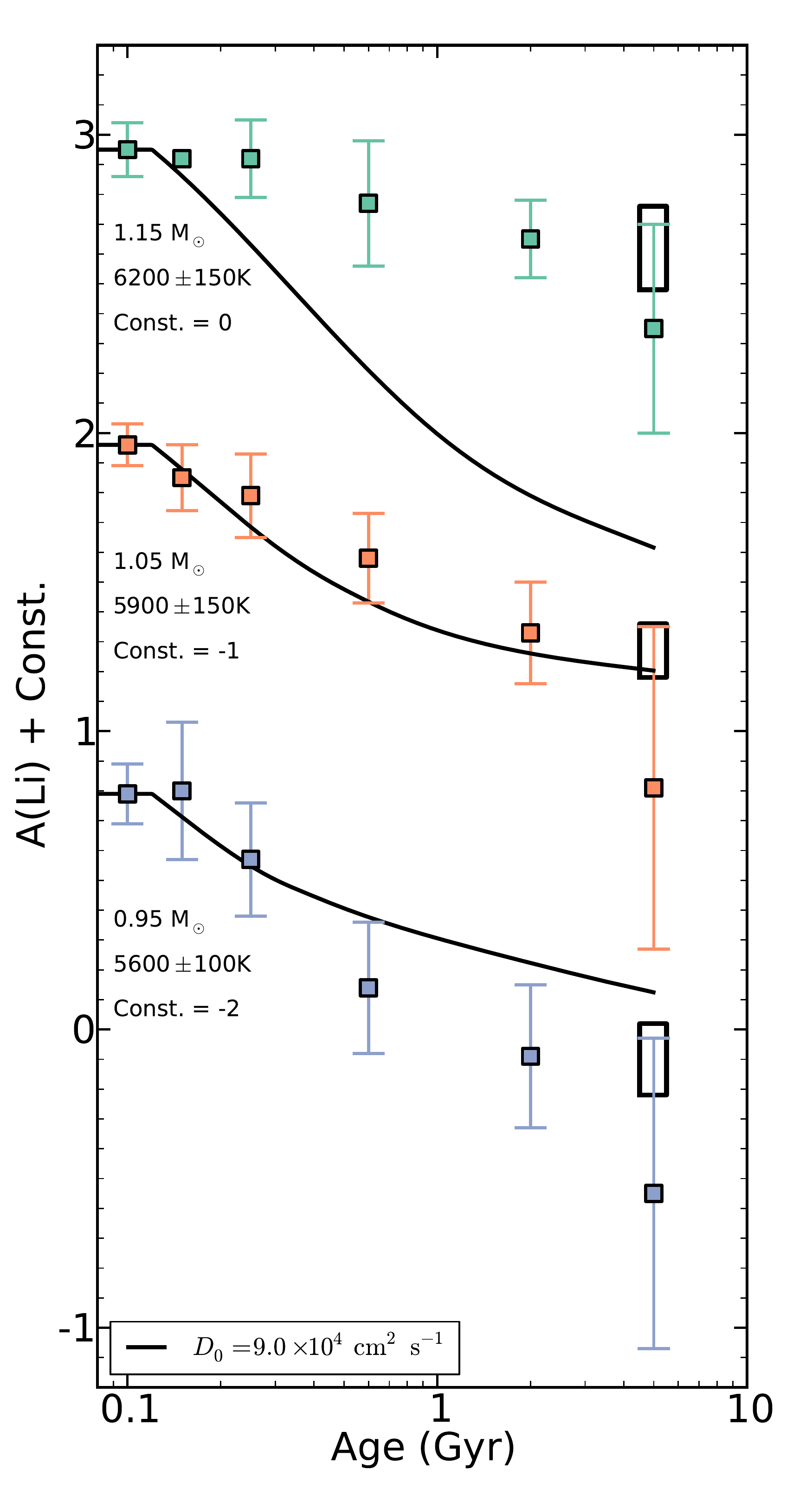}
\caption{Li abundance as a function of age compared to fixed-\do\ hybrid models in three different \teff\ bins: $6200 \pm 150$~K (green), $5900 \pm 150$~K (orange), and $5600 \pm 100$~K (blue), with constant offsets applied in order to display on a single plot. Black lines represent, from top to bottom, 1.15\msun, 1.05\msun, and 0.95\msun\ hybrid models, with the mixing efficiency \fc\ and constant diffusion coefficient \do\ as determined by the solar mass calibration in $\S$\ref{sec4.1.2} and \ref{sec4.2.1}. These models slightly under-predict Li depletion in the low mass bin, slightly over-predict Li depletion in the middle mass bin, and massively over-predict Li depletion in the high mass bin.}
\label{fig6}
\end{centering}
\end{figure}

\begin{figure}
\begin{centering}
\includegraphics[width=3.5in]{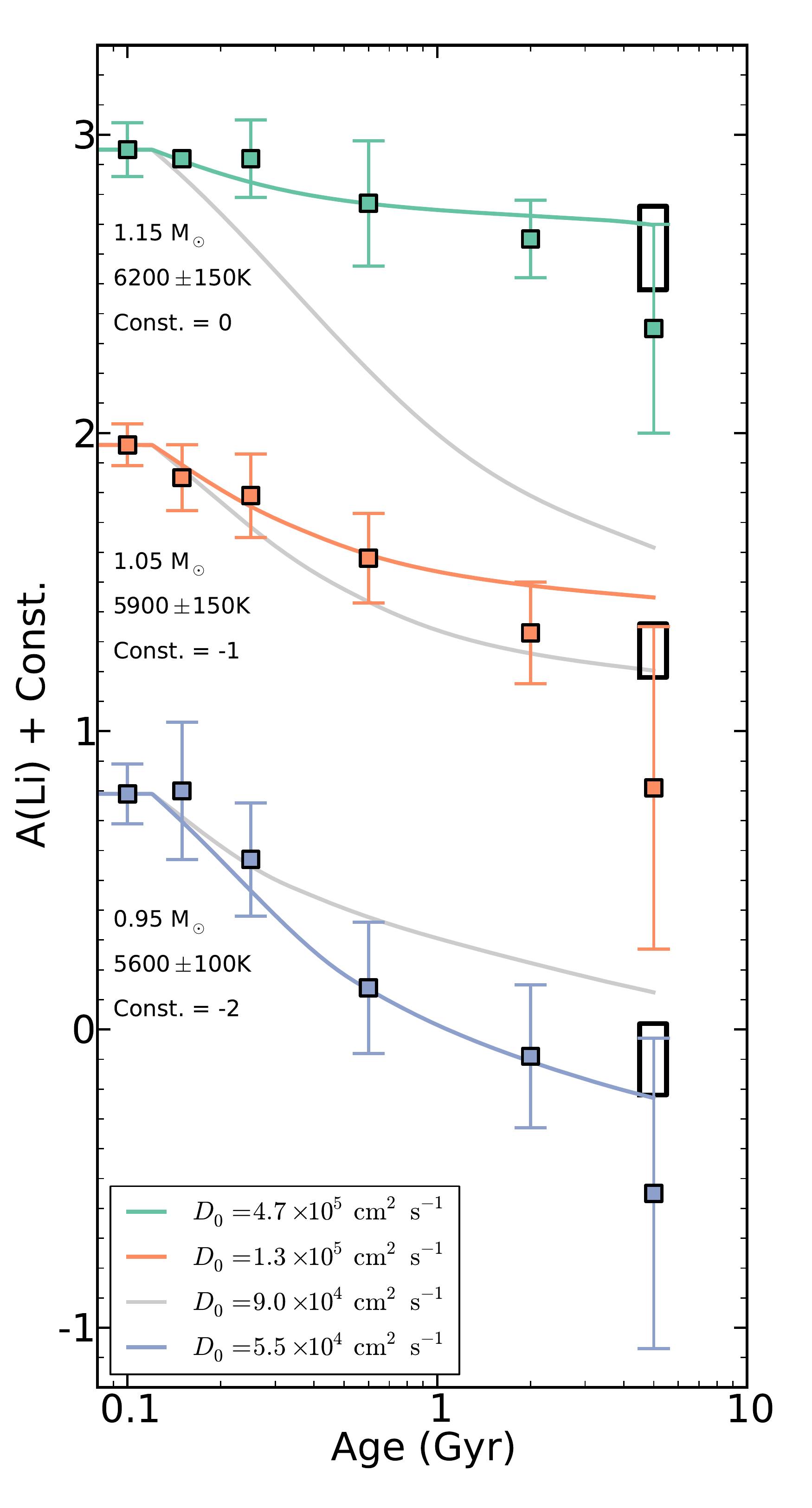}
\caption{Li abundance as a function of age compared to variable-\do\ hybrid models, in the same three \teff\ bins as Fig. \ref{fig6}. Grey lines re-plot the \do\ = $9 \times 10^4$~\cms\ models from Fig. \ref{fig6}, and the green, orange, and blue lines plot the hybrid models with the value of \do\ which best predicts the Li data. Each model uses the mixing efficiency \fc\ which was fixed by the solar mass case. Higher mass stars require systematically stronger angular momentum transport in order to predict the observed Li pattern. This demonstrates a mass dependence in the rate of core-envelope recoupling.}
\label{fig7}
\end{centering}
\end{figure}

\subsubsection{Mass dependence of \do}\label{sec4.2.2}

With \fc\ fixed by the solar \teff\ Li data, we can now make predictions for other masses. Namely, we compare our models to the average abundances of the three \teff\ bins considered by \citetalias{sestito2005}: $6200 \pm 150$~K, $5900 \pm 150$~K, and $5600 \pm 100$~K. We approximate these temperature bins respectively with models of mass $M = 1.15$\msun, $M = 1.05$\msun, and $M = 0.95$\msun. Adopting the same treatment for the pre-main sequence as in the solar mass case, we compute evolutionary models at these masses and compare them to the data in Fig. \ref{fig6}. Constant offsets of $-1$ and $-2$~dex have been applied to the 5900~K and 5600~K bins respectively. The models show reasonable agreement with the Li evolution for 5900~K and 5600~K (orange and blue), though compared to the mean values they are somewhat over-depleted and under-depleted, respectively. However, for the highest mass bin (green), the agreement is extremely poor. The models substantially over-deplete compared to the data at all ages, strongly ruling out moderately coupled models as an accurate description of 1.15\msun\ stars. Once again, M67 appears to defy the trend established by the younger clusters: its mean is significantly below the implied trend, its dispersion is far larger than any younger cluster, and its top-half distribution (black rectangles) appears to follow the trend implied by the other data.

We then allowed the constant diffusion coefficient, and thus the core-envelope coupling timescale, to vary for each mass individually, and solved for the \do\ which reproduced the 600~Myr Li bin. These models appear in Fig. \ref{fig7}. Each colored line shows the preferred \do\ for each mass bin, and the grey lines show the respective models from the previous figure. The best-fit \do\ values reveal a strong mass trend in the required strength of the additional angular momentum transport mechanism, with increasingly large transport strengths required to reproduce the mixing signatures of higher mass stars (see the inset key). The preferred models agree reasonably well with the upper half of the M67 data in each bin, in accordance with the conclusion from the solar mass bin. In this context, variable core-envelope recoupling rates provides a natural explanation for the different magnitudes of main sequence Li depletion observed in the different temperature bins.

\subsection{The timescales of core-envelope recoupling}\label{sec4.3}

In $\S$\ref{sec4.2}, we derived the constant diffusion coefficients \do\ that best reproduced the main sequence Li depletion signal for several stellar masses. In this section, we calculate the core-envelope coupling timescales \tauce\ corresponding to each \do, and compare the results to previous work. Note that this correspondence is only approximate, as the physics used to determine angular momentum transport differs between the two-zone and hybrid models. To obtain these estimates, we use the following equations from \citet{denissenkov2010}.

\begin{figure*}[t]
\begin{centering}
\includegraphics[width=5.5in]{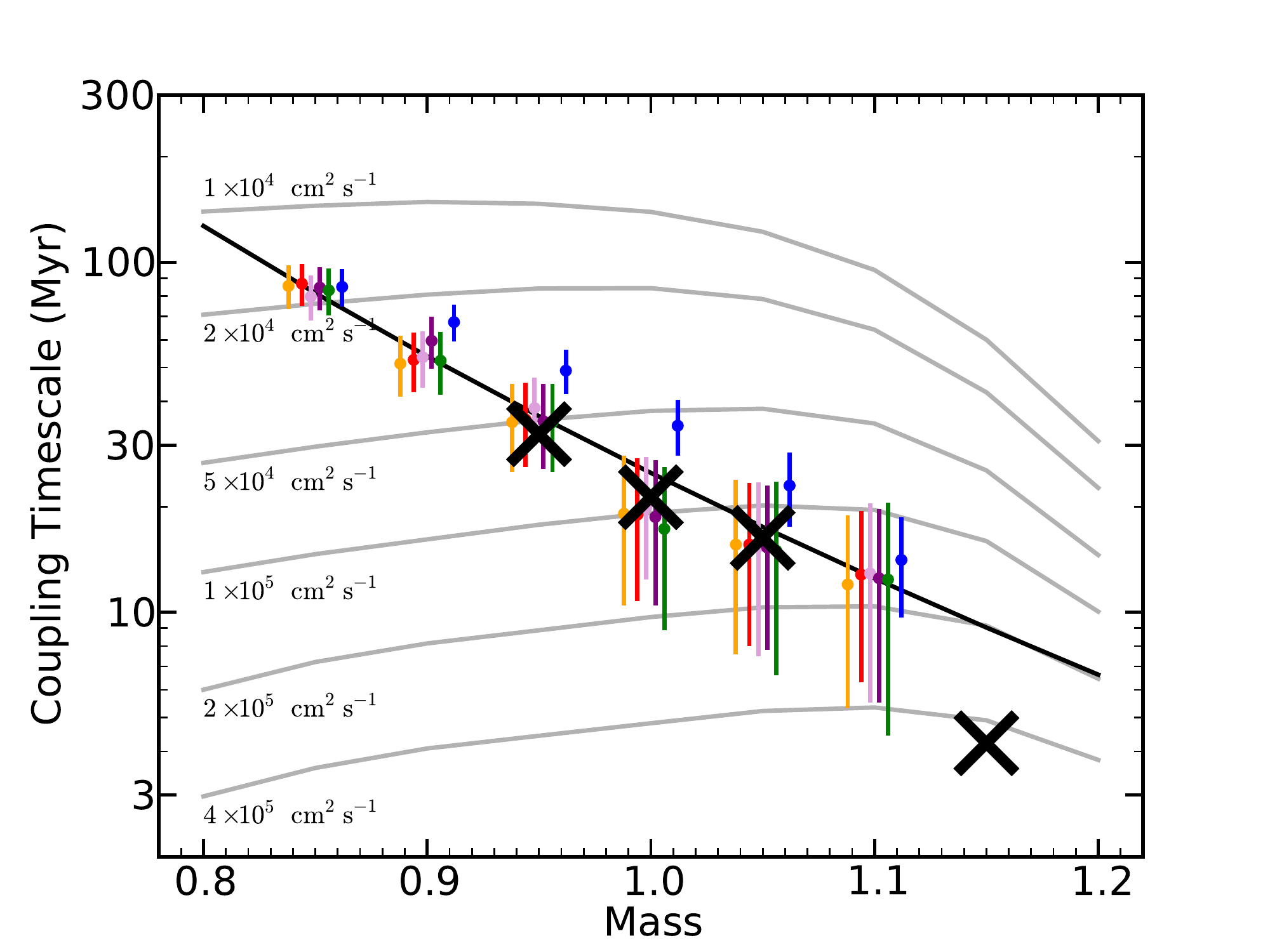}
\caption{The approximate timescales of core-envelope recoupling (\tauce) corresponding to our best-fit hybrid models. Grey lines show curves of constant \do, and the \tauce\ they induce as a function of mass. Our derived values of \do\ for 0.95\msun, 1.0\msun, 1.05\msun, and 1.15\msun\ appear as black X's. Our data are compared to the \tauce\ values derived from open cluster rotation by \citet{lanzafame2015}, represented by the colored dots. The black line is a power-law fit to their points. Agreement is overall excellent, confirming Li as a strong tracer of core-envelope recoupling.}
\label{fig8}
\end{centering}
\end{figure*}

\begin{equation}
\frac{dJ_{\rm c}}{dt} = - \frac{(\Delta J)_{\rm max}}{\tau_{\rm CE}} + \frac{2}{3}r_{\rm bce}^2 \Omega_{\rm e} \dot{M}_{\rm bce},
\end{equation}

\begin{equation}
\frac{dJ_{\rm e}}{dt} = \frac{(\Delta J)_{\rm max}}{\tau_{\rm CE}} - \frac{2}{3}r_{\rm bce}^2 \Omega_{\rm e} \dot{M}_{\rm bce} + \dot{J}_{\rm tot},
\end{equation}

where $J_{\rm e}$ and $J_{\rm c}$ represent the total angular momentum of the envelope and core, $r_{\rm bce}$ represents the depth of the convection zone, $\Omega_{\rm e}$ is the rotation rate of the envelope, $\dot{M}_{\rm bce}$ is the rate of growth of the radiative core, and $(\Delta J)_{\rm max}$ is set by,

\begin{equation}
(\Delta J)_{\rm max} = \frac{I_{\rm c} I_{\rm e}}{I_{\rm c} + I_{\rm e}} (\Omega_{\rm c} - \Omega_{\rm e}),
\end{equation}

where $I$ stands for moment of inertia. Each of these variables, except for \tauce, is an output of our evolutionary models. We extract these values from our model runs, and determine the average \tauce\ during the main sequence for a variety of masses and constant diffusion coefficients \do. The results appear in Fig. \ref{fig8}. The grey lines show curves of constant \do, and the core-envelope coupling timescale they produce as a function of mass. The timescales tend to fall off towards higher mass, due to the declining angular momentum content of the convective envelope as stars approach the Kraft break. The black X's show the \tauce\ values corresponding to the best-fit \do\ for each mass bin from $\S$\ref{sec4.2.2}. From these points, we see that the Li data indicates quicker core-envelope recoupling in more massive stars. The mass trend is quite strong in this range, with the favored \tauce's for 0.95, 1.05, and 1.15\msun\ equal to $\sim 32$, $16$, and $4$~Myr, respectively.

We compare these results to the findings of \citet{lanzafame2015}. They used rotation rates from open clusters to derive \tauce, so our mixing-derived coupling timescales are an interesting comparison for theirs. Their results appear as the colored dots\footnote{The dots have been shifted in mass slightly for increased visibility, but are measured at 0.05\msun\ intervals}. Each color represents different modeling choices (see their table 3), and the black line represents their power-law fit to the data ($\tau_{\rm CE} \propto M^{-7.28}$). Our results agree remarkably well, coinciding closely with the majority of the colored points from 0.95--1.05\msun. They did not consider 1.15\msun, but extrapolating from their power-law fit and assuming similar sized error bars as their 1.10\msun\ value shows probable $\sim 1 \sigma$ agreement. A power-law fit to our points finds $\tau_{\rm CE} \propto M^{-9.1 \pm 1.8}$, somewhat steeper than the value reported by \citet{lanzafame2015}, but driven entirely by the highest mass point for which they report no data. All in all, we find good agreement with rotation-derived core-envelope coupling timescales, confirming Li depletion as a powerful alternative constraint on internal angular momentum transport.

Our measurements indicate an extremely rapid fall off in \tauce\ above 1.1\msun. This is unlikely to be an artifact of our main sequence calibration procedure: a calibration age earlier than 125~Myr would only serve to increase the required strength of \tauce. A later calibration age would decrease the required strength, but is unlikely as stellar activity, the presumptive mechanism which our calibration corrects for, has little effect on rotational mixing. The error bars on our Li calibration point at 600~Myr are rather large, but the weaker coupling value of $\sim 2 \times 10^5$~\cms, required to match the \citet{lanzafame2015} extrapolation, produces too much Li depletion by 250~Myr to accommodate the earlier data points. A possible physical explanation explanation is the rapid disappearance of the convection zone as stars approach the Kraft break, coupled with the increases strength of gravity waves towards higher masses \citep[e.g.][]{talon2003}. The first effect substantially culls the amount of angular momentum contained in the envelope, and the second increasing the rate of angular momentum transport, so that recoupling occurs quite quickly. However, we note that our adopted wind law begins to break down around this mass \citep[e.g.][]{vansaders2013}, so definite quantitative results should be taken with caution. This result could be tested by measuring the core-envelope coupling in $\gtrsim$1.15\msun\ stars through spot modulation studies.

While a mass-dependent background source of angular momentum flux, along with hydrodynamics, can reproduce well the \teff-dependent Li depletion pattern of open clusters, the source of this transport remains unknown. As mentioned in $\S$\ref{sec1}, gravity waves are a strong candidate, as they efficiently extract angular momentum from the deep core without directly mixing stellar material. Numerical calculations find timescales of a few $10^7$~years for recoupling in solar mass stars \citep{zahn1997,kumar1997}, and a increasing momentum flux towards higher masses, up to the Kraft break \citep{talon2003}, consistent with our derived normalization and trend. By contrast, magnetic fields do not predict a mass dependence in the rate of recoupling \citep[e.g.][]{charbonneau1993,barnes1999}, in disagreement with our conclusions. 

\section{Discussion}\label{sec5}

\begin{figure}[t]
\begin{centering}
\includegraphics[width=3.5in]{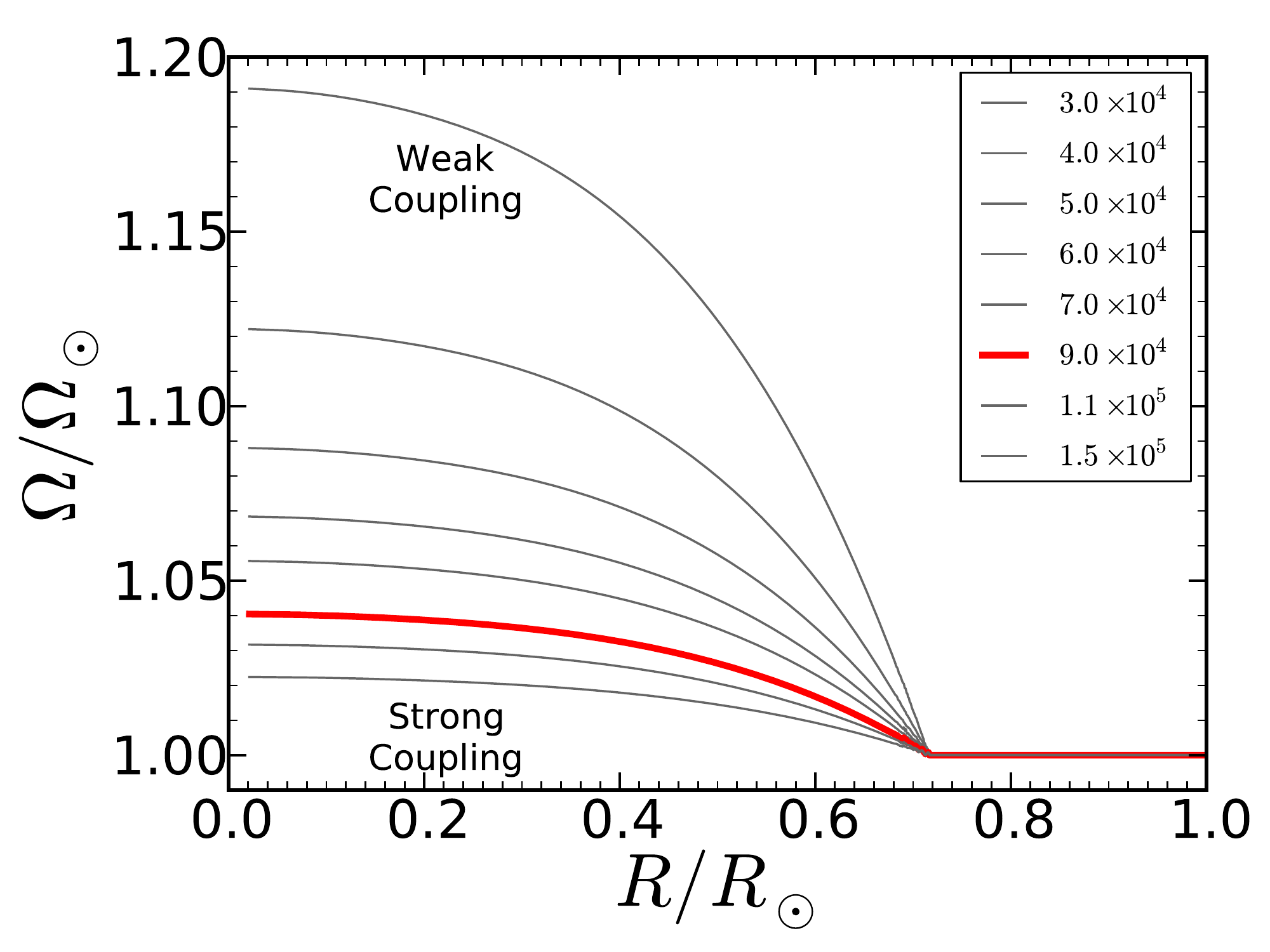}
\caption{The internal rotation profile of several solar-mass hybrid models at the age of the Sun. Our preferred \do\ of $9 \times 10^4$~\cms\ shows a 4\% core-envelope differential rotation, within the observational limits derived for the Sun of $\lesssim 10$\%. This suggests our models predict strong enough angular momentum transport to explain the solar profile.}
\label{fig9}
\end{centering}
\end{figure}

\subsection{The solar rotation profile}\label{sec5.2}

An additional empirical constraint on the efficiency of core-envelope coupling is the internal rotation profile of the Sun. Numerous authors have used helioseismic inversions to infer the solar rotation profile, and have found it to be extraordinarily flat within the radiative core, down to $\sim 0.3$\rsun\ \citep[for an extensive review, see][]{howe2009}. These findings strongly contradict the predictions of ``hydro'' models, which retain substantial internal differential rotation at the solar age (see $\S$\ref{sec3.1}). However, our hybrid models, due to their enhanced transport, dramatically suppress the degree of late-age differential rotation. While such a simplified model cannot be expected to reproduce all the details of the solar rotation profile, it can provide reasonable limits on the average rate of angular momentum needed to supplement hydrodynamic transport in the Sun.

Fig. \ref{fig9} shows the internal rotation profiles extracted for a range of \do\ values. From top to bottom, the strength of \do\ increases from $3 \times 10^4$~\cms\ to $1.5 \times 10^5$~\cms, with our preferred solar case highlighted in red. The convective region, which extends down to $\sim 0.72$~\rsun, is solid body by design, but below the tachocline we see the onset of radial differential rotation. Differential rotation below the surface convection zone is measured to be of order 10\% in the Sun \citep[e.g.][]{couvidat2003}, which we take as a reasonable overall bound on differential rotation in our models. This is achieved for \do\ $\gtrsim 5 \times 10^{4}$~\cms, corresponding to a core-envelope coupling timescales \tauce\ $\sim$ 35~Myr; this represents the weakest possible average rate of transport required to match the Sun. Our best-fit solar value of \do\ $\sim 9 \times 10^4$~\cms\ produces coupling above this level, showing that our derived rates of angular momentum transport are consistent with the Sun.

\subsection{Why is M67 so different?}\label{sec5.3}

We noted in $\S$\ref{sec4} that the oldest cluster in our Li sample, M67, seemed to deviate from the trend established by the young clusters. Namely, its average abundance shows a steeper decline from the clusters at 2~Gyr compared to the prior trend, and its dispersion is larger by 1.5-4$\times$ compared to the other points, depending on the temperature bin. In this section, we speculate on the cause of this difference, focusing in particular on how to reconcile the 2~Gyr old clusters with the 4~Gyr M67. We see three possible explanations: 1) incompleteness of the Li samples, 2) additional mixing setting in after $\sim 2$~Gyr, 3) differences in the initial rotation distributions of open clusters.

First, it is possible that the 2~Gyr clusters are as Li-dispersed as M67, but observations have failed to capture this dispersion due to small number statistics. This explanation has been previously explored by \citet{sestito2004}, who computed the likelihood of the relatively narrow \teff-Li relation of NGC 752 (2~Gyr) and the highly dispersed \teff-Li relation of M67 (4~Gyr) being drawn from the same distribution at two different ages. They demonstrated at the 4$\sigma$ level that NGC~752 does not possess as many stars depleted in Li by $>$5$\times$ as M67, thus concluding that the dispersions of the two clusters are intrinsically different. Furthermore, the three Li-studied 2~Gyr clusters (NGC~752, IC~4651 and NGC~3680) show no statistical evidence for significant differences in their Li patterns \citep{sestito2004,somers2014}, pinpointing M67 as the outlier. This leaves two explanations: either the dispersion appears after 2~Gyr, or sets in before 2~Gyr in some, but not all, open clusters.

If one believes that the open clusters form a true evolutionary sequence (an assumption we have made in this paper), then in order to explain M67, a Li depletion mechanism must kick after a few Gyr which is not only stronger than the depletion occurring on the early main sequence, but is highly variable between stars. Most depletionary processes are insufficiently strong and do not vary from star to star at the required level, such as mass loss \citep{swenson1992} and microscopic diffusion \citep{richer1993}. On the other hand, rotational mixing can in principle vary between stars with different rotations rates. However, rotational mixing tends to decrease in efficiency with advancing age, due to the gradual spindown of stars. Moreover, the converging rotation rates of the core and envelope suppress internal shears, further inhibiting mixing. Finally, the spread in rotation rates necessary to drive unequal Li depletion in different cluster members is only present in young ($\lesssim 500$~Myr) clusters, and is certainly gone by 2.5~Gyr \citep{meibom2015}, let alone the age of M67 \citep[e.g.][]{barnes2016}, due to the progressive convergence of rotation rates at fixed mass. Given the patent failure of the best candidate mechanism, the existence of a physical process which created the M~67 Li spread after 2~Gyr would be quite surprising, and would clearly represent new physics. While we cannot rule out this possibility, we consider it an unlikely explanation.

\begin{figure}[t]
\begin{centering}
\includegraphics[width=3.0in]{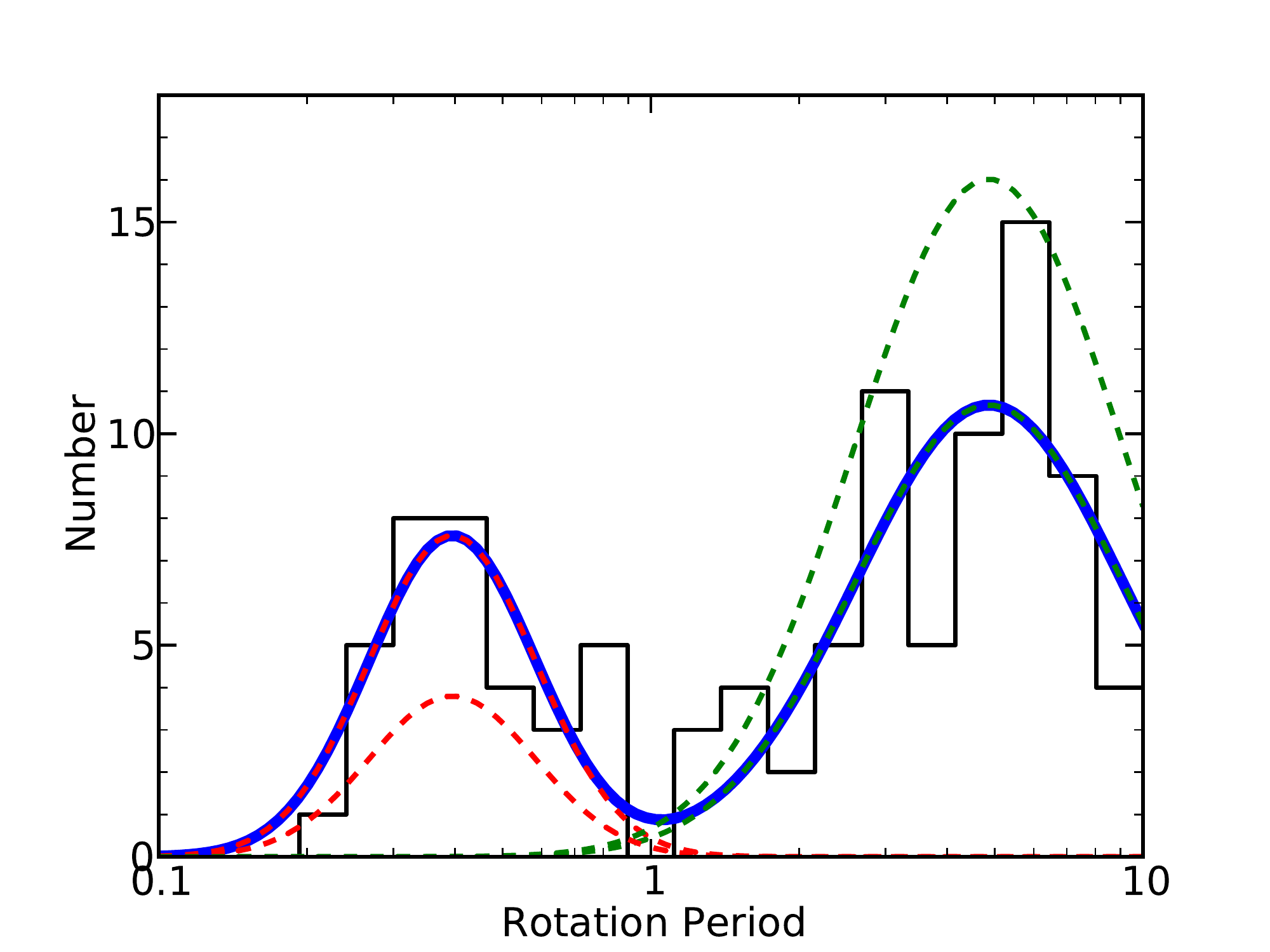}
\caption{The black histogram shows rotation rates of 0.9--1.1\msun\ h~Persei cluster members, as derived by \citet{moraux2013}. The blue solid line represents a double-Gaussian fit to the double-peaked profile, and the over-laid red and green dashed lines show the individual Gaussians. The lower red dashed line, and the upper green dashed line, show an example of a re-normalized initial distribution with a lower rapid rotator fraction than h~Per.}
\label{fig10}
\end{centering}
\end{figure}

\begin{figure*}[t]
\begin{centering}
\includegraphics[width=6.0in]{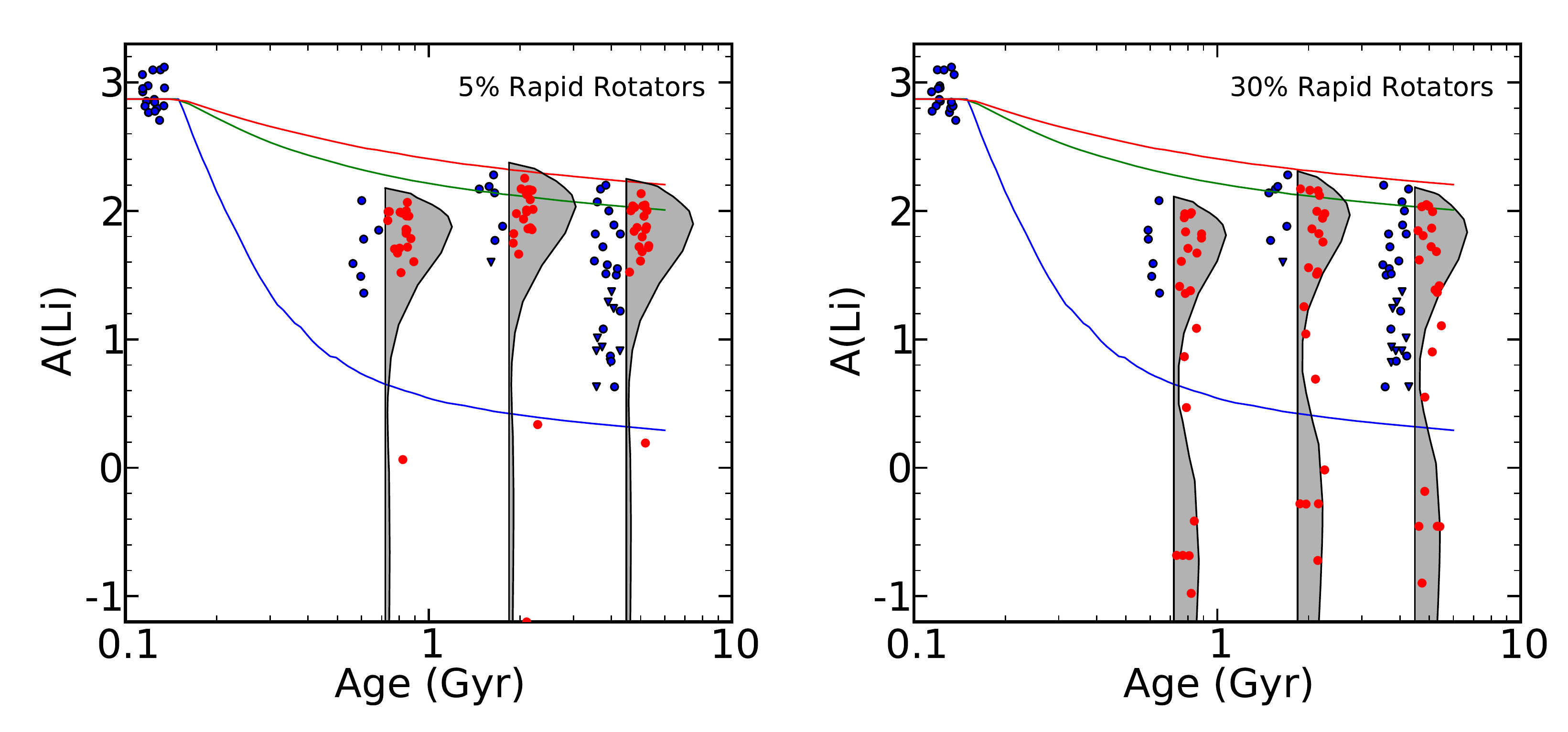}
\caption{Forward modeled Li abundances for synthetic clusters with different initial rotation distributions. \textit{Left:} Blue points represent 0.95-1.05\msun\ members of, from left to right, the Hyades, NGC~752, and M~67, randomly offset in age for visibility. Circles reflect detections, and triangles represent upper limits. The grey histograms show the relative fractions of expected stars at different abundances for a population with 5\% rapid rotators (red points are one such realization). Both the Hyades and NGC~752 are consistent with few rapid rotators, but M~67 is not. The lines reflect Li depletion histories for three different initial rotation rates. \textit{Right:} Same as the left, but for a synthetic population with a 30\% rapid rotator fraction. This distribution is a better fit to the wide dispersion and numerous upper limits in M~67, but strongly contradicts the two younger clusters. This may be a hint that the clusters were born with different rotation distributions.}
\label{fig11}
\end{centering}
\end{figure*}

The only remaining possibility that occurs to us is that M67 had an \textit{intrinsically different mixing history} than the other clusters in our sample. For instance, if M67 was born with far more rapid rotators than the other clusters, a larger spread would be a natural prediction of the rotational mixing framework. Inhomogeneity in initial rotation distributions has been recently demonstrated by \citet{coker2016}, who found that the 13~Myr h~Persei is rich in rapid rotators compared to the 1~Myr Orion Nebula Cluster. The cause is unclear, but may relate to the number density of proto-stars in the early cluster environment, and the subsequent influence of dynamical interactions between cluster members and the circum-stellar disks of their siblings. Such interactions could truncate the disk lifetime, thus terminating its extraction of angular momentum from its host star, resulting in greater rotation rates. This bounty of rapid rotators then undergo severe Li depletion on the early main sequence, creating a spread which persists for the lifetime of the cluster. 

To illustrate this possibility, we generated synthetic open clusters with different initial rotation distributions. First, we extracted rotation rates from h~Persei members between 0.9 and 1.1\msun\ from the catalog of \citet{moraux2013}. Their rotation distribution appears as a black histogram in Fig. \ref{fig10}. This distribution is clearly double peaked, which we identify as rapid and slow rotating components. We fit the distribution with a double Gaussian to determine the peak and width of the two components. We then varied the normalization of the two peaks in order to generate a rapid-rotator-poor and rapid-rotator-rich cluster distributions, with, respectively, 5\% and 30\% of members drawn from the fast component. Finally, we forward modeled these synthetic clusters to the main sequence, normalizing on the Pleiades abundance as before, and compared the predictions to stars between 0.95--1.05\msun\ in the Li catalog of M67 \citep{pace2012}. For comparison, we show two additional open clusters, whose abundances have been analyzed by \citet{castro2016} in a fashion consistent with these M67 data: the Hyades and NGC 752. 

The comparison appears in Fig. \ref{fig11}. The cluster data appear in blue, with detections indicated with circles and upper limits with triangles. One immediately notices the striking dispersion in M67, accentuated by the large number of upper limits. The grey histograms alongside each cluster represent the number of expected members, as a function of A(Li), for the rapid-rotator-poor distribution on the left, and the rapid-rotator-rich distribution on the right. For the left panel, the main peak is strongly focused around the higher abundance stars, corresponding to the location of the totality of the members of the Hyades and NGC~752, with the possible exception of the lone upper limit at 2~Gyr. The red points represent a single sampling of the distribution, revealing a tight clustering at the high end, with a few rapid rotators which have been \textit{heavily} depleted. This distribution is a good match to the two younger clusters, but is strongly inconsistent with M67, with its ample spread and numerous upper limits. By contrast, the rapid-rotator-rich histogram in the right panel is double-peaked, permitting a large number of Li-rich and Li-poor members. The random sampling shows a far more even sprinkling down to very low abundances, due to its strong rapid rotator component. While clearly inconsistent with the younger two clusters, the large number of heavily depleted stars presents a reasonable facsimile of M67. 

While merely a illustrative case with a simplistic initialization, this exercise makes clear the extent to which initial rotation conditions influence the subsequent rotational mixing pattern in a given open cluster. It also makes a salient prediction: large Li dispersions caused by rotational mixing \textit{must set in early}, when stars still spin rapidly. Thus, if this scenario holds, not all clusters of similar age and metallicity should host equivalent Li patterns. If the number of rapid rotators in the cluster is indeed related to the density of the star-forming environment, then a correlation between cluster density and Li spread should be latent. An attractive target for testing this theory is the dense open cluster M37, whose rotation pattern has been well studied but who lacks any Li measurements to date. A larger Li dispersion in this cluster compared the the similar-age Hyades and Praesepe would be powerful confirmation that Li spreads are a fossil remnant of the rotational properties of stars in their earliest days. 

\section{Summary and Conclusions}\label{sec6}

In order to investigate the impact of core-envelope coupling on rotational mixing, we have computed rotating stellar models whose angular momentum transport is supplemented by a background source of transport, parameterized by our models as a constant diffusion coefficient (\do); we refer to these as ``hybrid models''. To compare these to data, we devise a calibration methodology which accounts for several previously confounding aspects of the evolution of Li abundances in open clusters. These include the impact of the early disk-locking phase on the angular momentum evolution of young stars, the shortcomings of using the Sun as the sole calibration point for rotational models, and the anomalous rates of Li destruction on the pre-main sequence, which we argue are suppressed by a cooling of the convection zone by strong magnetic activity. By employing a post-disk-locking cluster for choosing initial rotation rates, a main sequence initialization point for Li abundances, and an ensemble of open clusters to constrain the angular momentum evolution and mixing efficiency, we present a modernized approach to studying Li abundances on the main sequence and beyond.

Our hybrid models behave similar to models which employ only hydrodynamic transport at early ages, and similar to solid-body models at late ages. As a consequence, while the traditional limiting cases fail to predict both the early and late evolution of the distribution of rotation rates in open clusters, hybrid models provide an excellent match for the observed pattern across the entire main sequence. Solar-mass models with a core-envelope coupling timescale $\sim$ 21~Myr provide a good description for the median evolution of the open cluster rotation pattern, proving that additional sources of transport are at work in solar-type, main-sequence stars. We also find a suggestion of a rotation dependence in the rate of this transport, such that faster stars drive stronger coupling.

Using open cluster Li abundances, we find that the required strength of \do\ is a strong function of mass, with stronger angular momentum transport required to explain the mixing patterns of higher mass stars. This implies a strong mass dependence in the rate of core-envelope recoupling, with timescales of $\sim 32$, $16$, and $4$~Myr favored respectively for 0.95, 1.05, and 1.15\msun. This dependence, derived entirely from mixing data, agrees well with the mass dependence revealed by rotation data. Our results shows a similar mass dependence as that predicted by gravity wave angular momentum transport, and can account for the near-solid-body rotation profile of the Sun. Finally, we show that M67 has a much lower median Li abundance, and a far larger scatter, than any other open cluster we have considered. We suggest that this results from differences in the initial distribution of rotation rates in the formative years of the cluster, perhaps associated with the density of the star forming region. The dense open cluster M37 is a valuable test bed for this theory, were its Li properties to be studied.

These results, along with findings in our previous work \citep{somers2015a,somers2015b}, provide a complete story for the evolution of Li abundances in solar-type stars. During the pre-main sequence, Li depletion is reduced in rapidly-rotating objects by the profound effects of magnetic activity on the structure of stellar interiors. This suppresses the dispersion which would otherwise develop due to pre-main sequence rotational mixing. Once on the main sequence, the convection zone retreats and direct destruction of Li in the convection zones ceases. However, the strong shears which develop at the onset of the main sequence, and the progressive spindown of the surface due to magnetic torques, leads to a depletion of the surface abundance through rotational mixing. As the main sequence progresses, the core and envelope recouple on a timescale governed by both their mass and rotation rate, gradually reducing the rate of depletion, and plateauing to a fixed abundance at a late age. This picture makes a specific prediction about the Li abundances of old open clusters -- their abundances will reflect a slowing of Li depletion at late ages.

We thank the anonymous referee for their helpful comments, and Jonathan Irwin and Jerome Bouvier for kindly providing us with the $\alpha$ Persei rotation data tables used for their 2009 review. MHP acknowledges support from NASA ATP grant NNX15AF13G, and GS + MHP acknowledge support from NSF grant \#1411685. The content of the paper was improved greatly by numerous conversations with the stars research group in the Ohio State astronomy department.


\end{document}